\documentclass[runningheads]{llncs}

\usepackage{hyperref}
\usepackage[T1]{fontenc}
\usepackage[utf8]{inputenc}
\usepackage{xspace}
\usepackage{graphicx}
\usepackage{listings, listings-rust}
\usepackage{subcaption}
\usepackage{xcolor}
\usepackage{xspace}
\usepackage{float}
\usepackage{lstautogobble}
\usepackage{color}
\usepackage{amsmath}
\usepackage{amssymb}
\usepackage{stmaryrd}
\usepackage{proof}
\usepackage{multicol}
\usepackage{mathpartir}
\usepackage{todonotes}
\usepackage{cleveref}
\usepackage{mathpartir}
\usepackage{wrapfig}
\usepackage{mathpartir}
\usepackage{amsmath, amsfonts, amssymb}
\usepackage{stmaryrd}
\usepackage[most]{tcolorbox}
\usepackage{transparent}
\usetikzlibrary{shapes,calc,arrows, arrows.meta,backgrounds}
\usepackage[misc]{ifsym}
\usepackage{orcidlink}

\definecolor{bluekeywords}{rgb}{0.13, 0.13, 1}
\definecolor{greentypes}{rgb}{0, 0.5, 0}
\definecolor{inferedgreentypes}{rgb}{1.0, 0.2, 0}
\definecolor{orangecomments}{rgb}{1, 0.5, 0.1}
\definecolor{redstrings}{RGB}{171, 114, 2}
\definecolor{graynumbers}{rgb}{0.5, 0.5, 0.5}
\definecolor{goldcomments}{rgb}{0.6, 0.4, 0.08}

\lstdefinelanguage{Lola}{
  keywords=[0]{input, output, trigger, constant, import, spawn, eval, close, with, when},
  moredelim=**[is][\transparent{0.6}]{?}{?},
  moredelim=**[is][\color{greentypes}@]{@}{@},
  keywordstyle=[0]\bfseries\color{bluekeywords},
  keywords=[1]{if, then, else, aggregate, defaults, offset, by, or, to, sin, cos, abs, hold, over, using, over_instances},
  keywords=[2]{Variable, String, Int, Int64, UInt, UInt64, Bool, Float32, Float64, Float},
  keywordstyle=[2]\color{greentypes},
  sensitive=false,
  comment=[l]{//},
  morecomment=[s]{/*}{*/},
  morestring=[b]',
  morestring=[b]",
  literate={\\@}{@}1
}
\allowdisplaybreaks

\newif\ifbadgeavailable\newif\ifbadgefunctional\newif\ifbadgereusable
\badgeavailabletrue\badgefunctionalfalse\badgereusabletrue
\RequirePackage{graphicx}
\usepackage[firstpageonly=true,angle=0,vpos=.147\paperheight,hpos=.393\linewidth,vanchor=t,hanchor=l]{draftwatermark}
\SetWatermarkText{%
\raisebox{-3.92cm}
{\ifbadgeavailable\includegraphics[width=11mm]{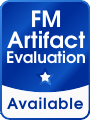}\hspace{.815\linewidth}\else\hspace{.905\linewidth}\fi%
\ifbadgefunctional\includegraphics[width=11mm]{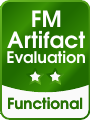}\else\ifbadgereusable\includegraphics[width=11mm]{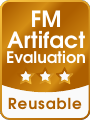}\fi\fi}}

\newcommand{\rtlola}{\textsc{RTLola}\xspace}
\newcommand{\stream}[1]{\texttt{\footnotesize #1}}

\newcommand{\spec}[2]{\href{#2}{#1}}

\definecolor{inputbg}{HTML}{E7FFD1}
\definecolor{outputbg}{HTML}{D2DAFF}
\definecolor{triggerbg}{HTML}{FFDCD1}
\tikzset{
  input/.style={
    shape=circle,
    draw=black,
    fill=inputbg
  },
  output/.style={
    shape=rectangle,
    draw=black,
    fill=outputbg,
    inner xsep=4pt,
    inner ysep=4pt
  },
  trigger/.style={
    shape=diamond,
    draw=black,
    fill=triggerbg
  },
  edge/.style={
  	->,
  	line width=1pt,
  }
}

\newtcolorbox{mybox}[1]{
    enhanced,
    overlay first={
      \draw ([xshift=1cm]frame.north west) -- (frame.north west);
      \draw ([yshift=-1cm]frame.north west) -- (frame.north west);
      \draw ([xshift=-1cm]frame.north east) -- (frame.north east);
      \draw ([yshift=-1cm]frame.north east) -- (frame.north east);
      \node at ($(frame.north west)!0.5!(frame.north east)$) {\textbf{For Experts: #1}};
    },
    overlay last={
      \draw ([yshift=1cm]frame.south west) -- (frame.south west);
      \draw ([xshift=1cm]frame.south west) -- (frame.south west);
      \draw ([yshift=1cm]frame.south east) -- (frame.south east);
      \draw ([xshift=-1cm]frame.south east) -- (frame.south east);
    },
    overlay unbroken={
      \draw ([xshift=1cm]frame.north west) -- (frame.north west);
      \draw ([yshift=-1cm]frame.north west) -- (frame.north west);
      \draw ([xshift=-1cm]frame.north east) -- (frame.north east);
      \draw ([yshift=-1cm]frame.north east) -- (frame.north east);
      \node at ($(frame.north west)!0.5!(frame.north east)$) {\textbf{For Experts: #1}};
      \draw ([yshift=1cm]frame.south west) -- (frame.south west);
      \draw ([xshift=1cm]frame.south west) -- (frame.south west);
      \draw ([yshift=1cm]frame.south east) -- (frame.south east);
      \draw ([xshift=-1cm]frame.south east) -- (frame.south east);
    },
    colback=white,
    frame hidden,
    breakable,
    before skip=15pt,
    left=1mm,
    right=1mm,
    bottom=1mm
}

\newenvironment{expertSection}[1]{
  \begin{mybox}{#1}}{\end{mybox}}
\begin{document}
\title{A Tutorial on Stream-based Monitoring\thanks{This work was partially supported by the Aviation Research Program LuFo of the German Federal Ministry for Economic Affairs and Energy as part of "Volocopter Sicherheitstechnologie zur robusten eVTOL Flugzustandsabsicherung durch formales Monitoring" (No.~20Q1963C), by the German Research Foundation (DFG) as part of TRR 248 (No.~389792660), and by the European Research Council (ERC) Grant HYPER (No.~101055412).}}
%
%
\author{Jan Baumeister\,\orcidlink{0000-0002-8891-7483},
Bernd Finkbeiner\,\orcidlink{0000-0002-4280-8441},
Florian Kohn$^\text{(\Letter)}$\orcidlink{0000-0001-9672-2398},\\
and Frederik Scheerer\,\orcidlink{0009-0007-8115-0359}
}
\authorrunning{J.~Baumeister, B.~Finkbeiner, F.~Kohn, F.~Scheerer}
%
\institute{CISPA Helmholtz Center for Information Security,\\Saarbrücken, Germany\newline
  \email{\{jan.baumeister, finkbeiner, florian.kohn,\\frederik.scheerer\}@cispa.de}}
%
%

\maketitle              

\begin{abstract}

Stream-based runtime monitoring frameworks are safety assurance tools that check the runtime behavior of a system against a formal specification.
This tutorial provides a hands-on introduction to RTLola, a real-time monitoring toolkit for cyber-physical systems and networks.
RTLola processes, evaluates, and aggregates streams of input data, such as sensor readings, and provides a real-time analysis in the form of comprehensive statistics and logical assessments of the system's health.
RTLola has been applied successfully in monitoring autonomous systems such as unmanned aircraft.
The tutorial guides the reader through the development of a stream-based specification for an autonomous drone observing other flying objects in its flight path.
Each tutorial section provides an intuitive introduction, highlighting useful language features and specification patterns, and gives a more in-depth explanation of technical details for the advanced reader.
Finally, we discuss how runtime monitors generated from RTLola specifications can be integrated into a variety of systems and discuss different monitoring applications.
\end{abstract}

\section{Introduction}

Runtime monitoring is an applied formal method that assures the safety
of a running system by evaluating its behavior against a formal
specification. In the stream-based approach, this
specification is given in terms of equations that relate input
streams, that contain raw data such as sensor readings, to output
streams that transform and aggregate the incoming information. The
values on the output streams are then checked against trigger
conditions that indicate faulty or dangerous situations.

This tutorial provides a comprehensive introduction to the \rtlola monitoring framework. \rtlola is
the real-time extension~\cite{DBLP:streamlab} of the \emph{Lola}
specification language~\cite{DBLP:Lola1}, which pioneered the
stream-based approach. \rtlola has been successfully applied to
cyber-physical systems such as (unmanned) aircraft. Major case studies
include the DLR ARTIS (Autonomous Rotorcraft Testbed for Intelligent
Systems) family of aircraft developed at the German Aerospace Center
(DLR)~\cite{DBLP:ClearedForTakeOff}, and the fully-electric aircraft
designed by Volocopter, a leading aircraft manufacturer of electric
multi-rotor helicopters~\cite{volostream}.

In the tutorial, we develop an \rtlola specification for a
real-world \emph{detect and avoid} problem from the aerospace
domain. We consider an autonomous drone flying in a shared
airspace. The task of the monitor is to detect aircraft in the
vicinity of the drone that might interfere with the drone's flight
path.  We will develop the specification in multiple steps, starting
with the simple case of a single non-moving object. Our final
specification will handle an apriori unbounded number of
independently moving entities. Along the way, we introduce the
relevant \rtlola concepts and some fundamental background.

There are two possible ways to read this tutorial. If this is the
reader's first encounter with \rtlola, we recommend focussing on the
development of the \emph{detect and avoid} example. The tutorial
starts in Section~\ref{sec:overview} with an overview of the
monitoring framework and the running example.
Afterwards, \Cref{sec:stream-basics}, \Cref{sec:timing},
\Cref{sec:lifecycle} and \Cref{sec:params} extend the specification
step-by-step, each section building up on the previous one.  Finally,
\Cref{sec:evaluation} explains how a monitor generated from a specification can be integrated into an existing system.

For readers interested in understanding the \rtlola approach
at a deeper technical level, the tutorial contains subsections with
additional background. These subsections are marked as \emph{for
  experts} to indicate that the subsections can safely be skipped at
first reading. In this spirit, the \emph{for experts} subsection of
\Cref{sec:overview} provides a comprehensive overview of the
various backends available in the \rtlola framework; in
\Cref{sec:stream-basics}, we discuss the static analysis of
\rtlola specifications. In \Cref{sec:timing}, we introduce the
type system, which is further refined in \Cref{sec:lifecycle}. In
\Cref{sec:params}, we discuss finer points of parameterized
specifications.

The tutorial is best experienced when following along in a browser window using the interactive \rtlola Playground~\cite{DBLP:playground}, which is available \href{https://rtlola.org/playground/tutorial}{\underline{online}}~\cite{PlaygroundTool}.
\Cref{sec:evaluation} briefly explains how this tutorial is integrated into the Playground.

\paragraph{Related Work.} There is a rich literature on runtime verification; we refer the reader to several introductory articles (cf. \cite{DBLP:journals/sttt/FalconeKRT21,DBLP:series/lncs/BartocciFFR18,DBLP:journals/jlp/LeuckerS09}).
A previous tutorial on runtime monitoring has focussed on writing monitors using aspect-oriented programming~\cite{DBLP:series/natosec/FalconeHR13}. Tutorials on stream-based monitoring have appeared as presentations at conferences (cf. \cite{DBLP:conf/rv/Torfah19,DBLP:conf/rv/Schwenger20}), but this is, to the best of our knowledge, the first tutorial paper that includes a hands-on development of a stream-based specification for a real-life application scenario. While the paper is based on the \rtlola framework~\cite{DBLP:streamlab}, the fundamental concepts apply in similar form to other stream-based monitoring approaches.
Notable other stream-based monitoring approaches, in addition to Lola~\cite{DBLP:Lola1} and its successor Lola2.0~\cite{DBLP:Lola2}, are the Tessla~\cite{DBLP:tessla} and Striver~\cite{DBLP:striver} tools.

\section{Overview}
\label{sec:overview}
Runtime monitoring checks the behavior of a system, such as cyber-physical system (CPS), at runtime.
The monitor receives input data from the system and analyses the data according to a formal specification.
At each step, the monitor outputs if the specification is violated such that the system or an operator of the system can initiate countermeasures to return the system to a safe state.
The next subsection introduces the general idea of stream-based monitoring followed by an introduction of the running example used in this tutorial.
\subsection*{Stream-based Monitoring}
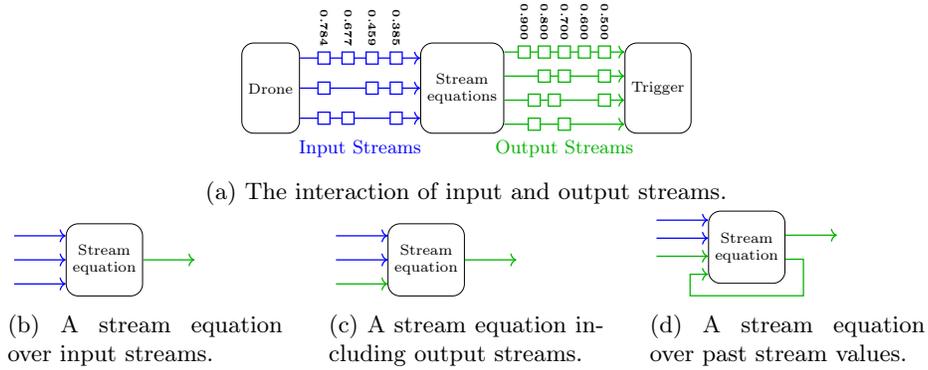
\begin{figure}[t]
	\centering
	\tikzset{
		n/.style={draw,minimum height=1.5cm,align=center,font=\scriptsize,rounded corners=2mm},
		in_s/.style={blue,thick,semithick},
		out_s/.style={green!70!black,semithick},
		timestamp/.style={font=\tiny\bfseries,rotate=-90},
		marker/.style={inner sep=1mm,draw,rectangle,fill=white},
	}
	\begin{subfigure}{\linewidth}
	\centering
	\scalebox{0.8}{
	\begin{tikzpicture}
		\node[n] (drone) {Drone};
		\node[n,right=2cm of drone] (equations) {Stream\\equations};
		\node[n,right=2cm of equations] (trigger) {Trigger};

		\foreach \shift in {-5mm,0,5mm}{
			\draw[->,in_s,transform canvas={yshift=\shift}] (drone) -- (equations);
		}
		\foreach \i/\ts in {1/0.784,2/0.677,3/0.459,4/0.385}{
			\node[timestamp] at ($(drone.east)!\i/5!(equations.west)+(0,10mm)$) {\ts};
		}
		\foreach \shift in {-6mm,-2mm,2mm,6mm}{
			\draw[->,out_s,transform canvas={yshift=\shift}] (equations) -- (trigger);
		}
		\foreach \i/\ts in {1/0.900,2/0.800,3/0.700,4/0.600,5/0.500}{
			\node[timestamp] at ($(equations.east)!\i/6!(trigger.west)+(0,11mm)$) {\ts};
		}
		\foreach \i in {1,2,4}{
			\node[in_s,marker] at ($(drone.east)!\i/5!(equations.west)+(0,-5mm)$) {};
		}
		\foreach \i in {1,3,4}{
			\node[in_s,marker] at ($(drone.east)!\i/5!(equations.west)$) {};
		}
		\foreach \i in {1,2,3,4}{
			\node[in_s,marker] at ($(drone.east)!\i/5!(equations.west)+(0,5mm)$) {};
		}
		\foreach \i in {1,2,3,4,5}{
			\node[out_s,marker] at ($(equations.east)!\i/6!(trigger.west)+(0,6mm)$) {};
		}
		\foreach \i in {2,3,5}{
			\node[out_s,marker] at ($(equations.east)!\i/6!(trigger.west)+(0,2mm)$) {};
		}
		\foreach \i in {1.5,2.5,5}{
			\node[out_s,marker] at ($(equations.east)!\i/6!(trigger.west)+(0,-2mm)$) {};
		}
		\foreach \i in {1.5,3}{
			\node[out_s,marker] at ($(equations.east)!\i/6!(trigger.west)+(0,-6mm)$) {};
		}
		\node[in_s] at ($(drone.east)!0.5!(equations.west)-(0,10mm)$) {Input Streams};
		\node[out_s] at ($(equations.east)!0.5!(trigger.west)-(0,10mm)$) {Output Streams};
	\end{tikzpicture}
	}	
	\caption{The interaction of input and output streams.}
	\label{fig:overview:stream}
	\end{subfigure}

	\tikzset{equations/.style={n,minimum height=1.2cm}}
	\def\offset{15mm}
	\begin{subfigure}{0.3\linewidth}
	\scalebox{0.8}{
	\begin{tikzpicture}
		\node[equations] (equations) {Stream\\equation};
		\foreach \shift in {-4mm,0,4mm}{
		\draw[<-,in_s,transform canvas={yshift=\shift}] (equations) -- ++(-\offset,0);
		}
		\draw[->,out_s] (equations) -- ++(\offset,0);
	\end{tikzpicture}
	}
	\caption{A stream equation over input streams.}
	\label{fig:overview:inputs}
	\end{subfigure}
	\hfill
	\begin{subfigure}{0.3\linewidth}
		\scalebox{0.8}{
	\begin{tikzpicture}
		\node[equations] (equations) {Stream\\equation};
		\draw[<-,in_s,transform canvas={yshift=4mm}] (equations) -- ++(-\offset,0);
		\draw[<-,in_s] (equations) -- ++(-\offset,0);
		\draw[<-,out_s,transform canvas={yshift=-4mm}] (equations) -- ++(-\offset,0);
		\draw[->,out_s] (equations) -- ++(\offset,0);
	\end{tikzpicture}}
	\caption{A stream equation including output streams.}
	\label{fig:overview:outputs}
	\end{subfigure}
	\hfill
	\begin{subfigure}{0.3\linewidth}
		\scalebox{0.8}{
	\begin{tikzpicture}
		\node[equations] (equations) {Stream\\equation};
		\draw[<-,in_s,transform canvas={yshift=4.5mm}] (equations) -- ++(-\offset,0);
		\draw[<-,in_s,transform canvas={yshift=1.5mm}] (equations) -- ++(-\offset,0);
		\draw[<-,out_s,transform canvas={yshift=-1.5mm}] (equations) -- ++(-\offset,0);
		\draw[->,out_s,transform canvas={yshift=2mm}] (equations) -- ++(\offset,0);
		\draw[->,out_s] ([yshift=-2mm]equations.east) -- ++(3mm,0) |- ([yshift=-2mm,xshift=-3mm]equations.south west) |- ([yshift=-4.5mm]equations.west);
	\end{tikzpicture}	
		}
	\caption{A stream equation over past stream values.}
	\label{fig:overview:past}
	\end{subfigure}
	\caption{An overview over stream-based runtime monitoring.}
	\label{fig:overview}
\end{figure}

Stream-based runtime monitoring interprets the system's input data as streams, i.e. a temporal discrete sequence over rich data values, transforms these input streams into output streams and expresses violations in the system's behavior.
\Cref{fig:overview:stream} illustrates this idea:
On the left side of the figure is the monitored system, which is in our case a drone.
This system emits a sequence of timed values, which are called input streams.
%
Stream equations then transform these input streams into output streams. 
Intuitively, stream equations are comparable to variable assignments in imperative programming languages.
Yet, instead of only being evaluated once, like variable assignments, they are applied at every stream position to filter incoming data, compare values from different streams, or express complex temporal properties.
Stream-based specification languages provide different stream access functions to reason about input streams (see \Cref{fig:overview:inputs}) and other output streams (see \Cref{fig:overview:outputs}).
Stream equations can also access past values of a stream (see \Cref{fig:overview:past}) to express temporal properties.
Finally, special boolean-valued output streams, called trigger streams, express violations in the system's behavior.

The next section introduces some syntax and semantics of stream equations in more detail, but first, we outline this tutorial's running example.
%
%
%
%

\subsection{Running Example}

 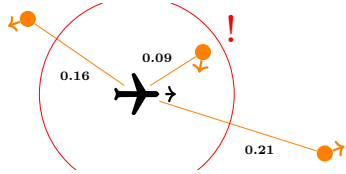
\begin{figure}[t]
	\centering	
	\scalebox{0.7}{
	\begin{tikzpicture}[intruder/.style={fill=orange,circle},trace/.style={orange}]
		 \node[rotate=-90] (plane) {\includegraphics[width=0.8cm]{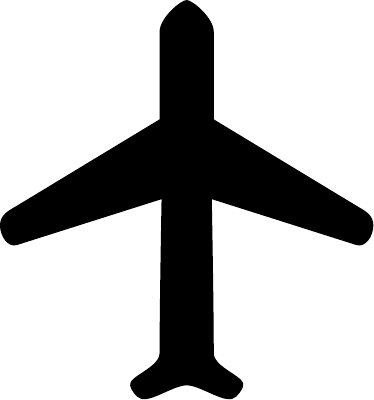}};
		 \draw[->,very thick] (plane.north) -- ++(2mm,0);
		 \node[intruder,above left=0.8cm and 2.5cm of plane] (intruder1) {};
		 \node[intruder,above right=1.2cm and 0.6cm of plane] (intruder2) {};
		 \node[intruder,below right=0.5cm and 4cm of plane] (intruder3) {};
		 \draw[trace, shorten <=2.9mm] (plane.center) -- node[below left,pos=0.4,text=black] {\scriptsize\textbf{0.16}} (intruder1);
		 \draw[trace, shorten <=3mm] (plane.center) -- node[pos=0.7,above left,text=black] {\scriptsize\textbf{0.09}} (intruder2) node[above right=0.8mm and 3mm,text=red] {\huge\textbf{!}};
		 \draw[trace, shorten <=4.5mm] (plane.center) -- node[pos=0.8,text=black,below left] {\scriptsize\textbf{0.21}} (intruder3);
		 \draw[->,very thick,trace] (intruder1) -- ++(200:4mm);
		 \draw[->,very thick,trace] (intruder2) -- ++(260:4mm);
		 \draw[->,very thick,trace] (intruder3) -- ++(20:4mm);
		 \clip ([yshift=3mm]intruder1) rectangle ([yshift=-3mm]intruder3);
		 \draw[red,thin] (plane) circle(1.85cm);
	\end{tikzpicture}}
	\caption{A drone monitoring surrounding vehicles.}
	\label{fig:overview:detection}
 \end{figure}

During this tutorial, we build a specification within the aerospace domain inspired by the \emph{Detect and Avoid} specification introduced by Baumeister et.al.~\cite{volostream}.
More concretely, we consider an autonomous drone flying in a shared airspace and describe the detection of surrounding vehicles that might interfere with the drone's flight path.
\Cref{fig:overview:detection} illustrates this property:
The plane in the figure's center represents the drone's current position and the rays around the plane, the distance to other participants in the airspace.
The exclamation mark indicates that this participant is close to the drone and is still approaching, so the drone should change the flight path to avoid a crash.
From now on, we will call these participants \emph{intruders} since they might interfere with the drone's flight path.

Algorithmically, the monitor should perform the following operations to monitor this scenario:
\begin{enumerate}
	\item \emph{Distance}: The distance to each intruder should be computed whenever the intruder or the drone moves.
	\item \emph{Closer}: To check if a specific intruder is approaching the drone, the monitor has to calculate whether the distance of the intruder is decreasing.
	\item \emph{Trigger}: If this approach continues over five seconds and the intruder is close to the drone, the monitor should notify the drone to initialize a counteraction.
	\item \emph{Stale}: The monitor should check if the intruder positions are regularly updated. Otherwise, it is declared out-of-range. 
\end{enumerate}
%
%

In this scenario, we expect the drone to provide the monitor with two sensor readings representing the input streams in our specification.
First, we assume that the system has a global navigation satellite system (GNSS) to provide the monitor with the drone's latitude and longitude.
Additionally, we assume that the system can detect other vehicles by sharing their position with other participants or by actively searching for intruders, e.g., with a radar, to get the latitude and longitude of the intruders.
As output, the monitor provides the current distance to each intruder and the trigger if an approaching intruder is nearby.


The following sections explain different specifications describing these requirements in \rtlola in a step-by-step fashion.
We start with an intruder with a fixed position, e.g., a tree.
Then, the specification is adjusted to handle a moving intruder, e.g., a single moving drone. 
Finally, we adapt the specification to detect more than one approaching vehicle.
%

\begin{figure}[t]
	\centering
	\includegraphics[width=0.7\textwidth]{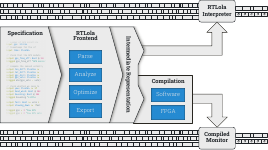}
	\caption{Overview of the \rtlola framework.}
	\label{fig:overview:framework}
\end{figure}

\begin{expertSection}{Integration \& Compilation}

The \rtlola specification language is embedded in an extensive framework to analyze and monitor specifications.
\Cref{fig:overview:framework} provides an overview of this framework.

It is divided into the frontend and several backends.
The frontend takes a specification file and produces an intermediate representation.
This representation contains an abstract syntax tree of the specification annotated with additional information relevant to the backends.
Additionally, the frontend optimizes the specification as presented in \cite{DBLP:optimizations}.
To verify the functional correctness of the specification, the intermediate representation can be inspected before executing the monitor, as shown in \cite{DBLP:verified-monitoring}.

All backends have online and offline monitoring capabilities, i.e., they can monitor a system at runtime or monitor a log of its execution.
\rtlola specifications can be executed in the software-based interpreter \cite{DBLP:streamlab} or compiled into the hardware description language VHDL~\cite{DBLP:FPGA} or imperative programming languages \cite{DBLP:rust-monitors}.
Providing both an interpretation and compilation ensures flexibility and efficiency:
An interpretation allows for easy debugging and quick development times of the specification as it can easily be adjusted and reevaluated.
A compilation, particularly a compilation to hardware, can provide highly optimized monitors that meet strict system requirements such as a low power consumption.
The framework's versatility was confirmed in industrial case studies~\cite{volostream} with aerospace partners.

In the \rtlola framework, the interpreter takes the specification as its intermediate representation and interprets it based on the incoming data from the system.
It provides an extensive API to integrate it into existing implementations.
Through the API, the interpreter can quickly adapt to different input data formats and sources.

The compilation takes the intermediate representation and produces executable code that implements a monitor for the given specification.
For software, it produces code in an imperative programming language such as Rust.
The hardware compiler produces VHDL code that can then be synthesized onto an FPGA.
The monitor implementation receives inputs through input wires, and the current stream values are stored on the corresponding output wires or variables.
After implementing the communication between the system and the monitor, it can be deployed with the system.
Although this approach is less flexible than the interpretation, the resulting monitor is highly efficient once built and integrated.

\end{expertSection}

\section{Stream-based Specifications}
\label{sec:stream-basics}
After \Cref{sec:overview} has introduced the general idea of streams and stream equations, this section presents the concrete syntax of \rtlola and gives the first concrete examples.
Output streams are declared using the \lstinline{output} keyword followed by a stream expression describing the computation of a stream value.
In comparison, input streams are declared using the \lstinline{input} keyword and do not require a stream expression as their value is given by the system under observation.
Trigger streams, special output streams with boolean stream expressions to convey violations to an operator, are defined using the \lstinline{trigger} keyword.

Stream expressions consist of common arithmetic and logical operators such as addition, subtraction, and conjunction.
Higher-level mathematical functions such as sine, cosine, or the square root can be enabled by importing the \lstinline{math} module.
To access past stream values at discrete positions \rtlola includes the \lstinline{offset(by: -n)} operator to access the n-th last value of a stream.
As the n-th last value of a stream might not exist, as the stream, for example, only has n-1 positions yet, an offset operation must be followed by a default value to choose in that case.

Consider the following simplified specification of the scenario explained in \Cref{sec:overview}.
We limit ourselves to a single non-moving intruder instead of multiple moving intruders and assume a synchronous timing model, i.e., all streams are evaluated when the input streams receive new values.
\begin{example}[\spec{A Static Intruder}{https://rtlola.cispa.de/playground/?spec=G2MBAJwHZZzH8sC82VlEXPw4iExZ09eoyytAMBZ4wF1QBUi_hcXltbyF_jO1i_W9s7EJTcg2U21KVKomJLWbQjfFV35NXC0l_BURxccciXLCPY77u4md2hPv2O7y5O9_3b4D4yXizHYj3TEtrOHZOrrrOUQyCrOMqMVvXxbFUr_QcI8lWbZ1Ewri5Kk8k5TZTll_4_rlbqgk_yejmx9rlqRvovxEFgMnCTkrg2jrLjHffg-4SE3dgoBuRJNBAA==&trace_name=c2ltcGxlX3RyYWNl&trace=G98wYJwJdqwVy-tUvAxje9Opej5VFfCk3OiK5cvvW_tP211qgSzFLriEtSVyK2Si_olQYW2r-8JA_ZmAf3-NyfHLPw6D-f4lOUFw7SLi0qX0cfrhHz79fL2--uPn_zvw_2p5_9bo2lV8-nj_am63kx0_8Fv8B6vyOavnL_9SR4F993Df-993o_c6UtTjLd1cDT7SWLyzt_RmPGxMnTF56dBzeHfAnGY4UPjNKY88PkLqQXjIHJcdeLPGo_Hk6VLMYnlP8NO3_wP4Tif29HZJed6coZcS41mgK4ELyvQrptbGoDvXBfNB8d4O6qFFkcabeiRsHw9d5z0_z8B83CdoYR_luFMkTx9P4wE2b4KX7fBQZijwBeOHSqTDOUTgvIt3J45c2t0LeYHTT_-zQfJO7RtQvl5vo6fQHXeffLfwE515ukdYaa3wojk88fMvRPl171ZCiFY71eOZGd6Rg6aqd_K9p9vioT0GV9cOoS-pqc7mja8f5-H1jk9iZg3FuDmI3ZnLi8Au3U73LNDK1_8BL-flxG9FzQQ3UnBQ99VsgoObKd_j3U3d9N29aLXgu-_-9x6mTw-ldih1yB0FG7s36FOK9cvnDL3XC_-HvTeHAaYcfPP-HCYcAO_1Q9Hjwg8V1CRz7vYUIb1eW-yLPIPrvKHiESMvmIX0OCJfX9GzzgXDmsF01lc4FwJbekErmigzO-Y7Xltt-urzwAz5LnN31-PGBHgWWs20AbLGWjvGgK7Lu2Sq4QBzZtJ7b2iyQ4zbdBXZ3Rcth7iU3CPRxcqH5VsEmVFBFxwDcHuoX7dNARJWR293XgOFeaXew8qIlTk3ebwGrpC8zlEPKZcEBI9v6VIb5SQuCfvBvL3DTqUV2KEvnLr1Lnp-a88AfcL1IiCLMnMtwbtZvGgf3u4BMQ_xEDxTVtVdL3xlxVk72AunAIvnQPMw2kfNoWiqZSWk0sBFF8bCN_3JlfL6XjqrFFHf06iHQY-7YVaLRYyMqXZYLiylId6b0U_PEvL6ChMdPDBzqxG-Pb0t8545eI7HY-5uda9w85hX9bbfPjphk7HDxltIzfNFABeztt-OE936xc-9fV2KVct1nc2377_rY0Uc3hNeLYyrjll4fAYdPzwpVk_actGa8p1Cjw8kRl-NGnh-lLCx3Rnl5m6i867Zzr0cPMzc7mPIxOZd49vM_QTXPfCFeGb7kHgOp6c3vSv8xF1PLyDNJ53FLK_UrLy7eyeAZajxmveasYrVnPDYeYEpPYTee3zDbCc8KEDmkDwrr6l5RJ1WlwY1rw-7axNEPHNZ4zzp2dDLm0XvPXf9et083s4OvTsVlp5d9WYGAHeOiXU8wryZ880hSsVTl53Tqajc8I4rUndaDt7DG3hecSX6wHm-3jl4I1D7-Hz69ngqToxHyqDGb66b7rjurqFH9q2eLLk85l0dM7Q2DC-2yEXOGwiv2Rz1cvAUbB3f_7CjIkyNhHTpi8ChsGfM9L2BQVDULlHCeD3RAznW-nykzFOVRYsJ7zDv3sOiRIe5d9otepiZMspAXTwwVp7lBh_ef2biTcSdIYrOnqvEhEPD5bpYTVHeSmlyerkL34OcF8Llh_3f3m5WTniPXK1Ps7fXh-47P-T5qo_xyzHVdltbm2uq3Zc9Q9rqk5vzKEWGDt7zFPYZ8zkYveI905hg7z2SuzIXWpxCOmxQ3XsvfE-D2NvnNxI9QUfztDp0lmV3EA3LN_c0nvNjpjqpmdVldudY3bjq4FWQ6FfeeQZ2bgQafbBdtiZm8pb5oO_9yZPBcs51-jRTiBj0msdYhCG-BTF3u5dD3rdLplvH-R9eSAznkPX9NdvRrRCaYH16uFHo8NS8-mlhvtu3-4DZJwd3xUjZoNYoc17Ox7fYe7PLSxmYt_Bw5cfI3wlmK8gRVct3TBY4VF5CtyM6qlJlJi7Mn7tv9-n4aKVnn1WeltTu8xE6kcjKFJviESdGfCXTW8I07JfGvM5uYYznvVfPx8f5vck-brVqMhG41MakViBbwnMxbq6WNW-WB4IPD7E-7v_eA1aHm3DQgHOLtyC3qTbQ6iSi0uGST6483mMfb6pBg-VT0Tcbcu_dG79dDoZ3yAKLRrHWe2yZlROXzSHxWtCu3upwx2EK2NNz8oLMSerwhG6Ps404LsdruZX7Rros9AJmng_kJMztdI_Eq7ZT7YH0vOROmZFPEu2u5Klm_544O29CedkyXreXorzObA8EwQdmkConK491lZd0RvXIibWa0TnkC6l3YwBgO9kMsOvZGb-Ehvlc3EZujrB6c4pe-wLKMXfHJ5djZZ5Fuki6r-HRvW2qwVGewg9o8RbTjoNrZ952EIqXaG4Jvsj3duLlSvGCrSrDZZUWddJBN_z0fuHmnd7Tg5jCe3k33EVCZwYT5Qvu4ZVtk9m5dugrro9xcncZvV4ul1zrvUOy1-TdHd6aUfkaYfgo-LWb4vQeLROaT_u_9zp5fsxTE9yntxqT4K58e1mQ8M72dtebr_d1Wco5HXIe3C7f05C5izcTvQSBezszrjW4d8GgBY7y2dsdPVoHnfaFswS8BOOBZzYCtsbo0U8XVt-LVroc2nlvb0I861n38jkk7vlbuGLB6WU7sWmn0XQWHFbeOTlSu9m93a6QfPIvRezxhkuZdTPdzKgNjHU8xFYkkze59-5aZThWnAPkT-mvAt_m3aW0AqNEwh47fHyLwFvx3XofRbln3oOu-yi9fmbKHYmyjsyBN8IliOsqevJa4QrZwpcZm7MPG5NH08E57x2eTriOVU3pXmlFJmqw2SX4wozfw94Y1OJgD18-v_-9tWLNvVkqIVXc4E2oymwm70mhizg8RGjLd0iePZ2uPn7Kpxvi2aTWHdGa8FjuImayhqW3l5b20vBV8XO8gFi5xAyZIczyuvTlUmMEZtju6WYz8aZzwPNunZhq-2hXDR7YqXf3xVyVJ8N5F4JuPZkJYAVdKkYeas28Cx9JnafO7g43Lz32onHkbF9dBoHR9OyEV8Xx7qCL3EcXG89Ygu7IW-F1sWpywuxqBlLpQHWhQJdUruwujGVgCa_4piZru_B6wmNxoz64eEWwcvXeFkYkREJe5VfccC-q5TT_A0YBa2yHcp5Gb2cFspxnDhJW7DHJGZRr-V6OA87dSqt_83qdweqF4XqUt90I1O2-eVqs2YV25q1pPvY7ldt9dXJ3ShmdduTjU8NVT6Gltm-t8Iuf29LXVT3weV_ZuX75C1PRqu8eZG5Pz8TtKsRp397j0c4YN8rkxaAV3pu-dYPoxLi-kbHeI5Fdr-eB6J6LR-zR0L5eyIfqzZNR36JvVH3d9Lpt1l3H4821Dwhpyh3kSO4lL2Ac_9hzozetTud8nZwjDuXdFTKbHYiGkZeuVwt_QmDsPN9LG-r5PXfYi-_x6Yw878tG1BPlMCQfNGb5rry7a3S9KfUyY90KsrUJPA11QfWc97B6s-wUwez1DtjOTaysBFIzR8q4LV8co528jqfLOtxNhsJRHgvnGQd76d69U7P4vQsnE2TTTeepnKndxy_535vRRujemcEt3-1kQ-CbbLFwsHJoY99ce7jMqbM6PqqTu4uOkAUfMSqfB57je03Ke1EJzjsh88pAXgAwsV5dh_c_7M2bV9lb9h1eMXtVAW53lWdOEVsMo9c320u-3uy7wIu_Pv_e7DOubx6oPN9gi120V3UznafBu30P3fA1ctksIqGpmAXPTfB4xzGnrM4JINg2N6lwutOutUcYEqO9B-NVXztq6gTucivUvlsv8u4FHsAgQDjdt8rkezqH8zwnHW7hB-KAWfEQzSNHC2Nj45s-RtVCzQGMfHR6d3PJrk76tqrRGByP6wq-cbpHlhyqp4a5y1rutVv-5m5zeSYl5FDKN-47xvuIw_oBLm69ngvz7PF7fLe2su3khlieq3kRDETOvpTjim-uz0Qo3HLmNaTduTDevnkYXMwddLT491Kpt9Mhs1tD0_OyZ42b11H46W7BuQdncr_QL293SFEDZTQPoxUPHGBCgXJfm_rVMFRbifYtnxWacybFW6g4ReHYCz1jW8aaTOkjmKUAM6GZ5HozwOxid-25VE_fvXV3ntkueJTN3TFuAvL0HqPlc3hNidnNqpNODq9t97EY57XMt2oZFg_RywzR7cs9VIvH413Y2aTpw3ZynE6c3N0TnQegZrAvppwFNsMgL2dnfPG1J8zI1fQW89J7gbatt6linycuNfagDp-O4qyLEz4f3IE5fqdpSajq2YHmBn_XHXvvbWHhrc716fzBvcdnTPEyeQJcO9tQJ60B4epQ_8M-T0frKdnc00J5HSDU0zPzAO4o9GlkNSSRICQWuu0L7StyWq7v3s4ijl-tHwent-AF1ALAXc33Ou_prRk6uXMOoY9dWgh2OrchMq9-syKju9DSa0mTD8HBTdapHlfawZtJM24f5QkWvR2-x7HdvFlh54IeW3l36_eaay2ty2aMMbVv3ccC7IYge7eYnZsX74ghb6fKkxddb-rDG-2Wr1t2HM6R6IgA1DwI71VCNLkiPtpSoLW8ofv8rHk38Tdv_ofBkR2rt5g87j42OVcClLsX48m0PF0LGGsXAsLIEPsWxDtd3ogBSiV4AmamrTipsANc_bSsYWQoUg5Tb3d3CNEzS8bXrBUXmeI5FACObIn0lTPe0dseiYcVcHcXjzpcrvOQnY4BrPGKh5VCxs3sPEFQ46fCs4-nG2f635uhMdYWXKqr2o0j8KvoyRjPrEbvQY-x867X08ZgRrd9956eL9hwt2g2ekWADTZXBKaOjCcPh8bhDvQwuTsRI7Sra8P1PT-KL11inpP6FXxx3x2jlbEq4r2jpNVrR9Aki1dw83KAiW6getc3gEBIs2eo3MZcg2emj6OdBy5sdXwStftUaZ6F8QVit0dXTxv0sxfE5Rtxj1tB59Y8_15xzHTyloMwj4yz_ggI7r45vNq7g_AwaWMmUweFKNXHzRy2vJY5no83WwjxnEyq3vmbZDTYbNBGzzuzyYp0fe2jb19iOC3YiF9FA8cbZEgpRi_jTGHCYF_kNTq4_N0FHGNLrKrSSasBKmCFdrsOvmf1LAJpavnh-SZpgY_z3tu1zt27IaNMQjnOssZI8dgljDts2gYv8D5Ee9Gd6n_gHqxRX-cUgOcDXdzMm9F7d8cK3FUIx7p3Mwbd4-7s7W2zf0O08ixbTlcBfYKQnK7QnEH2JDHuyXjOfVvCkvU83hLPO6MV9_o48XJB-MJCr1v0fLKHDLMNHw5Hu6MpyhakeguQ5ltgBGpOiT5eZx97c0a4fVmQtZH2-mZw6OzFuScXeZ0bCzq9zdyMj8RM5gJzwXY2KJ80SJ-ent_s7X1LQ3gB7RN1Xd5kFBAnvL19NMzcAdtG1IzV0qN7u8rEiHlt2RlGh68MHZ1APGRu4OUpfCmZSclnG4Qt-oGyVKzTaPRe3typBk-S5oDN9nIIAZBu20yqa7xdPr43mHcTMGjF66PJhd9MKT9Bu9p3c7FhnZOZboMxHDzf8765btDtQyI-tVTjhZ77imEzuZ33zpy2bofCA9vwpXeU_CxMSAkuPX49EpxhFssHiNhb31ys4AXJ4Pzkmnav83xhrdsu5xbA7pC6AgjnWkrUzawWe5hZzTh9XRk7fdnl3tsVUbn4wS9Tkn7rBxoWZHf97EoVOJFPbljaWDsLTicP0qqhPGFzjUNqt3OdLRfLTf3I14LU5oqe2xnDh3t6F8X6Gb_C182uUNPX9zovvH2r-rarK_Y0L3nad6Mn34lmMQaf3MLTbw9qjeL5qIUG2tftB-XTS8VpZiEVzXjQt8_4ufPw9lJUjDLfMiAvZHbqyRvroKlmXm5i8zl5EZzisaEbw2-q7p6-duDte7oOiUbUMY_LMGy5HAqm5Nu2I7FhzKIL-W0BBc486oUiy3sZvzdRjDA5vLEPv5OsxeL1GntAt9AZLag4c8OuSO6enNehHiZ2LZ62QcrDy2tMY3rXux1HEfrp1vx33K-H}]
\label{ex:static-intruder}
\begin{lstlisting}
import math
input lat: Float
input lon: Float
constant intruder_lat: Float := 249.301
constant intruder_lon: Float := 23.453

output distance?: Float? := sqrt((intruder_lat - lat)**2.0 + (intruder_lon - lon)**2.0)
output closer?: Bool? := distance.offset(by: -1).defaults(to: distance) >= distance 

trigger closer && distance < 0.1 "Too close to the intruder"
\end{lstlisting}
\end{example}

The two input streams \stream{lat} and \stream{lon} represent the measurements of the drone's GPS coordinates.
The constants below capture the static position of the non-moving intruder.
The output stream \stream{distance} keeps track of the Euclidian distance between the drone and the intruder.
To achieve this, it retrieves the current values of the input streams and uses the formula to compute the distance.
The type of the \stream{distance} stream is automatically inferred and can be omitted, denoted transparently in the example.
This inferred type follows from the floating point numbers given by the stream accesses and the underlying functions that operate on this type.
The output stream \stream{closer} captures the temporal property that the drone gets closer to the intruder.
For that, its stream expression compares the last value of the distance stream, expressed by the \lstinline|offset|-operator, with the current one.
Given that the stream expression consists of a comparison, the type of the \stream{closer} stream is inferred as a boolean automatically.
Finally, a trigger stream defines the condition when the distance is too close.
To make the trigger more precise, we require that the \stream{closer} stream also evaluates to \lstinline{true}.
Therefore, the trigger only activates if the drone moves towards the intruder.

\subsection{Semantics}
The semantics of an \rtlola specification is defined as a relation between input and output streams.
Intuitively, it compares every stream value at every timepoint with the computed value described by the stream expression.
The following definition gives the semantics~\cite{DBLP:Lola1,Schwenger_2022} for a subset of \rtlola to cover the general idea without focusing on concrete details.
\begin{definition}[Simplified \rtlola Semantics]
\label{def:semantics}
Let $\varphi$ be an \rtlola specification with input stream variables $i_1,...,i_m$ and output and trigger stream variables $s_1,...,s_{n}$. Let $\tau_1,...,\tau_m$ be streams of length $N$ of input values.
The tuple $\langle \sigma_1,...,\sigma_{n} \rangle$ of streams of length $N$ is called an evaluation model with respect to $\tau_1,...,\tau_m$ iff for each equation in $\varphi$ the following holds:
\[
\sigma_i(j) = val(e_i)(j) \quad\textit{for}\quad 0 \leq j \leq N
\]
where $e_i$ is the corresponding stream expression of $s_i$ and $val(e)(j)$ is for the expressions in our example defined as:
\begin{align*}
    val(c)(j) &= c\\
    val(i_t)(j) &= \tau_t(j)\\
    val(s_t)(j) &= \sigma_t(j)\\
    val(f(e_1,...,e_k))(j) &= f(val(e_1)(j),...,val(e_k)(j))\\
    val(e.\textit{offset}(\textit{by: }i).\textit{defaults}(\textit{to: } d))(j) &= 
        \begin{cases}
            val(e)(j+i) &\textit{for } 0 \leq j+i\\
            val(d)(j) &\text{otherwise}
        \end{cases}
\end{align*}
\end{definition}


\subsection{Evaluation Algorithm}
In contrast to imperative programs, the order of the equations in a stream-based specification does not influence the order of the evaluation.
They are generally evaluated simultaneously, yet accesses between streams imply dependencies between streams and therefore affect the order.
We represent these dependencies in a graph-based representation called the dependency graph.
An analysis of this graph then computes a correct order of the stream evaluation:
Every topological order of the dependency graph represents a valid order to process the streams during a single evaluation cycle.
\begin{definition}[Dependency Graph]
    Let $\phi$ be an \rtlola specification.
    The dependency graph of $\phi$ is a directed weighted multi-graph $G = \langle V, E \rangle$ with $V=\{i_1,...,i_m,s_1,...,s_n\}$.
    An edge $e=\langle s_i, s_k, w \rangle$ is in $E$ iff the expression of $s_i$ contains $s_k.\textit{offset}(\textit{by: }w)$ as a sub-expression.
    Analogously, edges with weight 0 are added for non-offset accesses.
\end{definition}
\begin{example}[Dependency Graph]
	The following graph describes the dependency graph for the specification in \Cref{ex:static-intruder}:\\
	\begin{center}
        \scalebox{0.8}{
	\begin{tikzpicture}
	\node[input] (lat) at (0,0) {lat};
    \node[input] (lon) at (0,1) {lon};
    \node[output] (dis) at (2,0.5) {distance};
    \node[output] (clo) at (5,1) {closer};
    \node[trigger] (t) at (7, 0.5) {T};

    \path [edge](dis) edge node[below] {0} (lat);
    \path [edge](dis) edge node[above] {0} (lon);
    \path [edge](clo) edge[bend right=15] node[above] {0} (dis);
    \path [edge](clo) edge node[below] {-1} (dis);
    \path [edge](t) edge node[above] {0} (clo);
    \path [edge](t) edge[bend left=15] node[below] {0} (dis);
	\end{tikzpicture}
        }
	\end{center}
\end{example}

Every node in the graph corresponds to a stream in the specification.
For a better illustration, we mark input streams green, output streams blue, and trigger streams red.
The input streams \lstinline|lat| and \lstinline|lon| do not have outgoing edges since input streams represent the input data and do not have a stream expression.
The output stream \lstinline|distance| accesses the current value of the \lstinline|lon| and \lstinline|lat| stream in the computation resulting in the 0-edge in the dependency graph.
The \lstinline|closer| stream accesses the current and last value of the \lstinline|distance|-stream, resulting in a zero and an offset edge.
Unlike input streams, trigger streams have no incoming edges since no stream expression can access these streams.
In our example, the outgoing edges of the trigger are the accesses to the \lstinline|distance| and \lstinline|closer| stream.

Using this graph, we can compute different evaluation orders for this specification ensuring that the monitor accesses the intended stream values:
The specification allows every order in which the \lstinline|distance|-stream is evaluated after the inputs, the \lstinline|closer|-stream after the \lstinline|distance|-stream, and at the end of the evaluation the trigger stream.

\begin{expertSection}{Static Analysis}
    This expert subsection presents two static analyses based on the dependency graphs that guarantee a safe evaluation of stream-based specifications.
The first analysis guarantees the existence of a unique evaluation model, whereas the second analysis computes an upper bound of the required memory.
The latter analysis can determine if the monitor can run in a resource-constrained environment before execution.
\paragraph{Well-formed Specifications}
With this analysis, we define a syntactic criterion to guarantee the existence of a unique evaluation model.
In general, we cannot find a unique evaluation model, if we cannot determine a order of the stream evaluation.
This problem corresponds to a cycle in the dependency graph.
First, consider the following examples of syntactically valid specifications and their dependency graph illustrating the problem.

\begin{example}[Ill-defined Specifications and their Dependency Graph]\\
\label{ex:ill-defined}
\begin{minipage}{0.25\linewidth}
    \begin{center}
    \begin{tabular}{c}
    \begin{lstlisting}
        output s := !t
        output t := s
    \end{lstlisting}
    \end{tabular}
    \end{center}
    \end{minipage}
\begin{minipage}{0.25\linewidth}
\begin{center}
\begin{tabular}{c}
\begin{lstlisting}
	output s := t
	output t := s
\end{lstlisting}
\end{tabular}
\end{center}
\end{minipage}
\begin{minipage}{0.5\linewidth}
\centering
\scalebox{0.8}{
\begin{tikzpicture}
    \node[output] (s) at (0,0) {s};
    \node[output] (t) at (2,0) {t};

    \path [edge](s) edge[bend right] node[below] {0} (t);
    \path [edge](t) edge[bend right] node[above] {0} (s);
\end{tikzpicture}
}
\end{minipage}
\end{example}
According to the semantics, the specification on the left has no evaluation model, since we cannot find an assignment for \lstinline|s| and \lstinline|t| that satisfies both equations.
Such specifications are not well-defined and should be rejected by the \rtlola framework.
In comparison, the specification on the right has multiple evaluation models as long as \lstinline|s| and \lstinline|t| are equal.
However, for this specification, we again cannot compute a valid evaluation order.
As a result, neither specification can be evaluated algorithmically.

Both specifications have the same graph with an edge from stream \lstinline|s| to stream \lstinline|t| and an edge in the other direction, resulting in a cycle.
Because of this cycle, we cannot determine a valid evaluation order and should reject the specification.
We can give a syntactic criterion for well-definedness called well-formedness based on the dependency graph of a specification~\cite{DBLP:Lola1}:
\begin{definition}[Well-formedness]
    \label{def:wellformedness}
	A specification is well-formed, iff for every cycle in its dependency graph, the accumulated edge weight of the cycle is not zero.
\end{definition}

As described in the definition, we do not forbid every cyclic behavior. 
The next example shows a valid specification containing a cycle in the dependency graph.
It sums all values of the input stream \lstinline|a|:
\begin{example}[Valid Cycle]
    \begin{lstlisting}
        input a : Int
        output sum := sum.offset(by: -1).defaults(to: 0) + a
    \end{lstlisting}
        
\end{example}
Here, the \lstinline|sum| uses the offset expression to access the past value of itself.
This access results in a cycle in the dependency graph, but the sum of this cycle is negative.
Intuitively, this is allowed since past values are already computed, and we can ignore these accesses when building the evaluation order.


\paragraph*{Static Memory Bounds}
Secondly, we can determine the amount of values that need to be kept in memory for every stream.
Again, this analysis is based on the dependency graph and allows giving static memory bounds for specifications.
\begin{definition}[Memory Bound]
	Let $G = \langle V, E \rangle$ be the dependency graph of the specification $\varphi$.
	For every stream $s$ in $\varphi$ its memory bound is determined as $max(\{-w \mid \langle o', s, w \rangle \in E\})$.
\end{definition}
Intuitively, the memory-bound of a stream is defined by the largest offset at which the stream is accessed.
All values before that bound are not required for further computation and can be discarded.
For the specification in \Cref{ex:static-intruder} the memory bound of all streams is zero except for the \stream{distance} stream.
Given that the stream is accessed with an offset of $-1$, the memory bound of this stream is one. 
The memory bound of the specification as a whole is then the sum over the memory bounds of all input and output streams.

\end{expertSection}
\section{Event-based \& Periodic Streams}
\label{sec:timing}
This section lifts the simplification of a static intruder to a moving intruder.
For this, consider the following naïve extension of \Cref{ex:static-intruder}, where the intruder position is given as additional input streams \stream{intruder\_lat} and \stream{intruder\_lon}.
\begin{example}[\spec{A Synchronous Moving Intruder}{https://rtlola.cispa.de/playground/?spec=G1ABABwHuUnzBm7jL0EJRJjT16gMUaKcscAH7oOq38Li8lreQv8ZtYvV78xDF4QsyEIXxaR2m0JP_dwiUHSP_GMShWeOUnLinlb0d5M6LZ5PXZ76_a_bd7inUe2ZhCvcK3dVKwC_fdX38SoJ1hSEYZnmgloj8shTeZYA3u0TvuIb8R_JGOdXJEWAmNOV-89XVfYW0JU2ROgFrnHdHxrytBAAkyMRc4oJ&trace_name=c2ltcGxlX3RyYWNl&trace=G5gtEZW6EkBngm2jceuWNYTa2IfhJVvYm9MW5d_Yt5fNUIDSJnOTo9MsX2V1_K9Vzna32kRFclwyclw8QqFxWBQVx3ELR1EehDxu43CuZ957vzfpnUQo1jj8n1ZTLeGp5VN7Nd31BzY5QVQXg9NExKuat379Gvb5x6__aYOPU36nX1-_fT35ob-c5-fJPzD8D43um4adpN6cf7pvjJspdpBH_YbfMa5z2fHLb7-RT6b67vF-_1--f99RfT8xvVfY655u6ew11JPCxYO5UWYRKI_HGWLfForZuyNxBgR69T5UJ52eaLeQgbSnUQHPTvgUPgcpE8PRPTPP5ud7pN75rJ7fLKkusxf6oSKCITs2OJSjvBJVRqgU16FWYOW9ARsolkXzoYHN6VPRFe_lBaBppM_0MDnx8c5rB306QTOcfVi-nUKHaiFTb4kcaktZ4bgG8W4zg81auTO30i2zFTyTUNKd2wdSuV5v1s9Bd5p5zt3Az8ri-Z4Yl8RjvjWOz0rIzHuT17kbm4u1HtRPJ1NhBs6ym1rvnHvPN-Vjylpe02RFJy1FfYkOuXMNj-9mzmZRQ294COQMcPvWVMpTdC6m4hU6Q769jLB5YwpYHuzlQZ3XqLs8uIvqPd0dqvzu3no81DvhLniP6PNj5QHlQhp4OWH3wD5vOT59WShzbcolcw9HkKhA7_C-WMQKJN_rVQpS5rGGuru4dHoqK2UyidW3fAte8SBvIEHeEkP7CSZfX9mLLwVXjZYoJq1kb0VOlaGsMNYLDKJ3ahgd5ZoLGC35bnF319OEmQyGHmMCgGPEkyw_6evobhcVeBFx0W7vPSh_3_zsNxS5FiCH3L1TmBV824wzxq_Va_vejPMey-W-XN7E73NJzL0WdTv-HFEQkXY7LpqZtx5BKLd370nscH41Pqh6w-UC3qTWy9OSmh4rKzDlnhAzh9m3hHl5xJnTTfLHpxv3Na4gu0X7Ng83RfM67ZaUYBd-M3hdulhX-T2Ow7UCl-4-XZep5OZXnPy41SjsM87oKzK38b0mnc5pcLPRpup689bF3B70gyn2vQKbDUmaQW6UUrPeszUSnMfo5o6DctVUodwKiWk7w17eJAAJr8IeckNAPZIPtzzvvawHWNYWgJ7sM5k6IfY9vNu5TSr0mddbkzuoFteKukPwsH18M0euZSs-Ll_kOD3veqtX1cKEpXdFmruUPCgHbgDzFWCBPUt-XnyzWr82cSHwQpxmdTMzt0JJlS8L4RGeJ-NQdutRbe76Qt76VhvToN_jjPfKe9u7-iQ2g8cFezZhwB48MAqrSVf2zopHgCD0tSM21-REUWJzz3APo59mVjsOtgiHqAeKzxzZEN8DxGuHEZ3JlREzhyUW91jkCTECr3DgzWaooW3hPULz3vrohisF2Pee9xoKPmpxYxifnt9U-150MU6n097dWFmzT3srv6nDeS1X3UWy6mEh7Rn3DchT8TBWCu8gKhMuc85U4dww93S_eUv5VrdvZIZuaXdfbk2ImCRvkF08zdu89OZpl3HcatLsrJ53farF43tWFAmcOIegbRn2uqPBBme0jsFp3gte7jD3IkILJ8pajVjqYXj1aQdHLlQ2j8-lfvZUwzb1DUo_PUqEfvM9siFfnmyOku_O97ASNlOVWG_gPq2e--zEaHzRXEcz0sjlVYtyptQM4D3cIexdd4p7e4Ohxc08rbR8eNfNzeKeRV7VWz2k2sdVMdc3z2_BtTy4AUZwX0_zkAx28ysXZ6XaiuTBcnDp2wbf4_n5oXcNftZM0FtKM40thhhd5ax4d_fOpKqVoI7vdRGXY5wZBB9oBmi6897QPYPc2XvzvBtbSvgux4erkyS1e2hG3ZLJvmVkPyLv6UE7xYNZUtrj7tvj6zY6sdlW8ICU8fruN5OYuqKwBy_EYKNeBDiGY797vj2jB1972neTWX_73tEMN2cOFDfA7eRK6uyX0G9f0HsvnbxeH_dmioEyAyEd3wtm3ANAEtnZTthZ6rx9XXc36CFd70yTnbrwDSblvo2dLLgNxuhLDk678ekEAZfDcb34c0fF-VyOL8wgteQ7aZvygcErr1LgIWG7LnhGNhlu8GE3mOHBTN5vPXxbDiit4u1y_fiMPgov17ss3jXleXo5F6MWNnySY7GRh-tsB3redRU4ufEzBeYSlENMVpXAe7A0PCt3Hfpb2KzE3Rf8pt--F3le6W5uKdVp3zVrim3CMMOt7K2zDzRf9_jYjMDn5VS_-x4HLlfbcJfJA2LhvRGGR4L0ewrOHaO33ofID4kdHXtb8_cly22ih-yITJVbU5CxFwJ9DwxIWZ4RK1o2wfpRQjx63tXRud5hy9h3nsnQGPfx3OoYfLqsXzXQyPL15HkveA749HU53LxriJnviHfvcdGuj7h3nq30CKBaL-iUwXAcjGZpw9cx2s3sWgOIMCYzmF4ACR1vGppvgHmiRtj9WoAlxtgf8e3TOw62ul3JYi71bhSqMNWkHF-sG3u7e37L9x6dfSum0nO-Nzez4-zqoihq50pN6Vzfqp5ZHyybbG87dY8p85A7G5_n58vOBVoquqRyPDljbq7UvMvjvlyNvz1tPZ02BrhuPfPAXOiOd073Am-5tXDH3JVTJJcnVJPAfUvyLcudcbNz90IITaSymKVhzkPB94IyuWC5LOFXvsUbUb33JO26VrFQ2sftqu_N3Xtv9Z5BxjtjN-92ie6hO1qC5wraTDKOBeLzB6_vNkrvTNPDcSdHvXl5sBUctsbz-FhAM-AaqnKWBd3g8rRoOt9FO77FDI7xMAFIZBCfQh0QdQfmFW75Hmeb690LDCIJ5n3DmUbmeLQPqQu-OiFWXnUXoNmDqbCP41MbEdg3WsmnP4eqcNb0oOFECnndTITOYDn2ljcZyW-M914aHJKiEBW8M8Mq-gyUFka_7tPGgmm9oYi7mfHRmTcjbaf67vf4ViKE406q1q8766vL0HfHKiN038bPnYSOKlw5YTfK7H4L6lX7YBZ-3Cl8YwZhOTk_HlR8fO6-5nlo2UzfzCMxzzrelbB3lo3H8ol3Lm8Qv7eHDgyf-aaLLDnyB7PtzvVktl8KKJCanKgIby-nN9h7mNFttbBuGGicp-V7Z0at6azUlqN32h3yFCQ9M4CBeVy_va0dgxvPzBh8ke22uZSbkc7SDh9UwqHOiL6BlaGuXC02pemXzpt5Pj1Z0Usurs4jCfekDX2WuOMw-DBiDoe9G_jaLU2LNwPPs1fOzMs4oYu3P73ImXdjEkBGqVs-8awluUrbG8EKk7fd6LrlMkTw3mvgUK9l3sPO09SrsMlMrxLysmZpZGF2uUifid0CKt6-2Yt8IkzgXZKOKAqLNTXylNmtKbWCcRsernFc3z6KenzcWO_53nvk-HjYKjzr9M5DRNStVKL4cMiOsBt1dN7WsxkGSWMNNvukl41Fg8IN34Cabj1Ljyu5rrIa6fF5n-6pT4cKbEdVz2UfZlk47CJ-myCE9XabZAcRzT1MtI1s51nnB-yJ1Ny-7EPyEHLu3UPejIDxHXfIYUM58XtqtcMn3s4edzOR1HT8xsc7QWHr7RPltnyu2kN2Y6N6Wr7dbE-eVnnrVgdRpublMq7lKHLOL7tvEZztQmd0esJ0LWR8fK2mTh8sujn9llq8HBneGdwSXKJiLFS3KHrOO2365DWZ6b3rCyVLPLjwUk0ldrU36JyIV09Rz4EK3u6dF-DD7npm7KBC2r5nYfCwcmLR5WZIOuy5m_f8kPLu7akPe8M7do2qJ4hSvtfL07y7w0abSXtbVFdgehQFPmrBrfdsxVNT79st4BO3KTYeA74shYl2pzeFmvd81OKCBdSJxEE27303e33AWlzJtwCjVXDkW8nvEJJgi50FOZOeg7xdhdFL2ZV7ODHu4bxim72lnI1m4MfXjecuh-BlUyFCmde9l7g9-Uizp2ANSrV2uc6bnXFyhHjxIrbD3c5rMHxvujV44tsyj2z5hhuuLTXJ4k3BJYLW4dYh69tsJQf7WsHbQAYXIYQ3S33-8kWbl8nM25fS7Ro3oh6Ue4PNaKzyllPXYaoqx_LZRx9k49f2Pb_ze360GPTtGyLFXp1QfZPIcl4ChaF9m4P8nrHz8_vSKBi88at1Fd_bd9AMr_BFS6wXnOOr2u5il3mgXHl9Wp3vbuHX29vb_IinOXFeYVw8XfYg2C9biwI8ydedeyctnkXIkqluMPFYbeaOu3M78e6Ob6JVy3VN6MnMa3w8v6c4oqHLfO-9Yl-e9rmxe02oHZxFAO_IOAfzeee46GJzdHyDw2OdprEOyEYvDaF5fmNEhGacm9sBxQymNzOJbzKvo8rZE9kUj2lH7xlU8t57sW99730cg1xp0sd9pGrsbjYttifQdMoDh69H7t1mFuu3geS9AZDGg3u3BFvyxIe5GfgpPopq-lYYkRmRpZ3nHMNBSNxsRwduRndWLpHflzp3erOXKKs-PM6bzhgBA8ya2ITwU55vsfveeuzbY4bvzWHF98eLp-NlJZ74hqlH8cnNW5D38DzNFr6WGd8ilDyYE-6h-9Bo-zQbTWcmuWIo9HaKZWjRNYoBoTqDc9ZWpDtzMx1z9zz5o_z0eHM6aKwevMCdjY723T5k9IgM3rGWX8N5Ilq8DRQi8zxF50o9ZrroLOC2APHpuInKLvZh7727lm-FeLNHOuK2fmvqzb47P7y5ecuGSc51otel4nsHZYUkuy8-DzB70hrpzANrUx5T610beyr09IYLT613k3my-AK8R1_nyWLbAagGthyP1Rs8570BMiJfkAqNgpubSmmWzNsYL8vElqD3iELeEBx1MG8DeR2vnzNRSc2dMrdALFMfZyOdIh0v-97x-czrWqV1OTw57K0mjjMeWN820UORvN5biumem3SKp1JGRaYb2fequATrLmdnCL3VIu8xZcrDYwK99R_mfe9NvDHuYeQYm2Y49WA8DTScgx37OnuNjzp2z3TE6eR8v7xvZym-jmavhF1rlg9B9WIW-55VfGtBxzXb-AaSLkHREdEe1fNBfImcnKjunUjd3ts7Bc8eAd8bWfb27PJZnil-wZdg7DJqikRNCivGqmzv1lrsBLbf3LZKRtyv3rxshpSm6YiAtBDDyvVeJnhE-qSHMRZVLTZ3e0j2HqZwLFYcOfeiV747janrKLe3DSYSUDvnTAGb2eLIZ-iztp7eM-PA9uSgmZm30X7fapLxLNkyFwUHpqdHb99YNkYXMxdjI22-3npwO0nUPofZdyswbbDAklFRKm-4j43vsg5zknxBTd0dD297-otQ}]
\label{ex:moving_intruder_sync}

\begin{minipage}{0.48\linewidth}
\begin{lstlisting}
import math
input lat: Float
input lon: Float
input intruder_lat: Float
input intruder_lon: Float

output distance :=
    sqrt((intruder_lat - lat)**2.0
        + (intruder_lon - lon)**2.0)
output closer := distance.offset(by: -1)
    .defaults(to: distance) >= distance 

trigger closer && distance < 0.1
    "Too close to the intruder"
\end{lstlisting}
\end{minipage}
\begin{minipage}{0.49\linewidth}
\def\distance{0.7}
    \centering
    \scalebox{0.85}{
\begin{tikzpicture}[circ/.style={draw,circle,fill=white,inner sep=1mm}]
    \node[anchor=east] (lat) {\texttt{lat}};
    \node[below=5mm of lat.east,anchor=east] (lon) {\texttt{lon}};
    \node[below=5mm of lon.east,anchor=east] (intruder_lat) {\texttt{intruder\_lat}};
    \node[below=5mm of intruder_lat.east,anchor=east] (intruder_lon) {\texttt{intruder\_lon}};
    \node[below=5mm of intruder_lon.east,anchor=east] (distance) {\texttt{distance}};
    \draw (lat.east) -- ++(3.1cm,0);
    \draw (lon.east) -- ++(3.1cm,0);
    \draw (intruder_lat.east) -- ++(3.1cm,0);
    \draw (intruder_lon.east) -- ++(3.1cm,0);
    \draw (distance.east) -- ++(3.1cm,0);
    \foreach \i in {1,2}{
        \foreach \j in {lat,lon,intruder_lat,intruder_lon,distance}{
            \node[circ] at ($(\j.east)+(\i*\distance,0)$) {};
        }
    };
    \foreach \i in {3}{
        \foreach \j in {lat,lon}{
            \node[circ] at ($(\j.east)+(\i*\distance,0)$) {};
        }
    };
    \foreach \i in {4}{
        \foreach \j in {intruder_lat,intruder_lon}{
            \node[circ] at ($(\j.east)+(\i*\distance,0)$) {};
        }
    };
    \coordinate (LL) at ([yshift=-3.5mm]distance.east);
    \draw ([yshift=2mm]lat.east) -- (LL);
    \draw[->] (LL) -- ++ (3.2cm,0) node[right] {$t[s]$};
    \foreach \i in {1,2,3,4}{
        \node at ($(LL)+(\i*\distance,-2mm)$) {\scriptsize$\i$};
    }
\end{tikzpicture}}
\end{minipage}
\end{example}

This specification is correct in a synchronous setting, i.e. a setting where all input streams receive a new value simultaneously.
Yet, in reality, the intruder and the drone are independent systems.
Hence, the measurements of the intruder's position might not be synchronous with the drone's position measurements.
In an asynchronous setting, input streams receive values independent of each other.
For this, every output stream and trigger is only evaluated if the input streams they (transitively) depend on receive a new value at the same time.
Consider the specification from \Cref{ex:moving_intruder_sync} in a synchronous and asynchronous setting as illustrated by the trace on the right.
First, all values are received synchronously and the \lstinline|distance| stream is computed.
It might seem correct, yet all output streams and trigger streams still transitively or directly depend on all input streams.
As a result, the specification is effectively synchronous again and behaves invalidly if not all inputs receive a new value at the same time, as depicted later in the trace.
As we can see, if the current position and the intruder position are not synchronized, the \lstinline{distance} stream is not updated as expected.


The next example depicts the corrected specification, where input streams are accessed asynchronously using \lstinline|hold|-accesses:
\begin{example}[\spec{A Moving Intruder}{https://rtlola.cispa.de/playground/?spec=GywCACwG7Ib6RYBpMY4SY1Guc_ofI4T0YM5OSUxU0ko1-Z1JuiTMZQ-VXoOkuCDtt7C4vIM3bQdkU28o_W_hO0APKB9ZgaT-dJtf1D9Z6TIQ-iK1iJK32QHKHb46U83r25-ggM-v9s9J1zRT--q0Eh8XtNLrFa6W5Vl4FEWukvmebJSxF8IU4qOTLD-DxttfStnsGIifp6rubBpyGT5MHz_vFbmOp5ZDoau_NmMxPGU4Tf_5kE-pPrWiBBQIUOuVtfNdMV4AJPreQ0vIPhWsX0FdphbUv17raIdI7FQUd4Fj54mczcFG_QtiFgA=&trace_name=c2ltcGxlX3RyYWNl&trace=G5gtEZW6EkBngm2jceuWNYTa2IfhJVvYm9MW5d_Yt5fNUIDSJnOTo9MsX2V1_K9Vzna32kRFclwyclw8QqFxWBQVx3ELR1EehDxu43CuZ957vzfpnUQo1jj8n1ZTLeGp5VN7Nd31BzY5QVQXg9NExKuat379Gvb5x6__aYOPU36nX1-_fT35ob-c5-fJPzD8D43um4adpN6cf7pvjJspdpBH_YbfMa5z2fHLb7-RT6b67vF-_1--f99RfT8xvVfY655u6ew11JPCxYO5UWYRKI_HGWLfForZuyNxBgR69T5UJ52eaLeQgbSnUQHPTvgUPgcpE8PRPTPP5ud7pN75rJ7fLKkusxf6oSKCITs2OJSjvBJVRqgU16FWYOW9ARsolkXzoYHN6VPRFe_lBaBppM_0MDnx8c5rB306QTOcfVi-nUKHaiFTb4kcaktZ4bgG8W4zg81auTO30i2zFTyTUNKd2wdSuV5v1s9Bd5p5zt3Az8ri-Z4Yl8RjvjWOz0rIzHuT17kbm4u1HtRPJ1NhBs6ym1rvnHvPN-Vjylpe02RFJy1FfYkOuXMNj-9mzmZRQ294COQMcPvWVMpTdC6m4hU6Q769jLB5YwpYHuzlQZ3XqLs8uIvqPd0dqvzu3no81DvhLniP6PNj5QHlQhp4OWH3wD5vOT59WShzbcolcw9HkKhA7_C-WMQKJN_rVQpS5rGGuru4dHoqK2UyidW3fAte8SBvIEHeEkP7CSZfX9mLLwVXjZYoJq1kb0VOlaGsMNYLDKJ3ahgd5ZoLGC35bnF319OEmQyGHmMCgGPEkyw_6evobhcVeBFx0W7vPSh_3_zsNxS5FiCH3L1TmBV824wzxq_Va_vejPMey-W-XN7E73NJzL0WdTv-HFEQkXY7LpqZtx5BKLd370nscH41Pqh6w-UC3qTWy9OSmh4rKzDlnhAzh9m3hHl5xJnTTfLHpxv3Na4gu0X7Ng83RfM67ZaUYBd-M3hdulhX-T2Ow7UCl-4-XZep5OZXnPy41SjsM87oKzK38b0mnc5pcLPRpup689bF3B70gyn2vQKbDUmaQW6UUrPeszUSnMfo5o6DctVUodwKiWk7w17eJAAJr8IeckNAPZIPtzzvvawHWNYWgJ7sM5k6IfY9vNu5TSr0mddbkzuoFteKukPwsH18M0euZSs-Ll_kOD3veqtX1cKEpXdFmruUPCgHbgDzFWCBPUt-XnyzWr82cSHwQpxmdTMzt0JJlS8L4RGeJ-NQdutRbe76Qt76VhvToN_jjPfKe9u7-iQ2g8cFezZhwB48MAqrSVf2zopHgCD0tSM21-REUWJzz3APo59mVjsOtgiHqAeKzxzZEN8DxGuHEZ3JlREzhyUW91jkCTECr3DgzWaooW3hPULz3vrohisF2Pee9xoKPmpxYxifnt9U-150MU6n097dWFmzT3srv6nDeS1X3UWy6mEh7Rn3DchT8TBWCu8gKhMuc85U4dww93S_eUv5VrdvZIZuaXdfbk2ImCRvkF08zdu89OZpl3HcatLsrJ53farF43tWFAmcOIegbRn2uqPBBme0jsFp3gte7jD3IkILJ8pajVjqYXj1aQdHLlQ2j8-lfvZUwzb1DUo_PUqEfvM9siFfnmyOku_O97ASNlOVWG_gPq2e--zEaHzRXEcz0sjlVYtyptQM4D3cIexdd4p7e4Ohxc08rbR8eNfNzeKeRV7VWz2k2sdVMdc3z2_BtTy4AUZwX0_zkAx28ysXZ6XaiuTBcnDp2wbf4_n5oXcNftZM0FtKM40thhhd5ax4d_fOpKqVoI7vdRGXY5wZBB9oBmi6897QPYPc2XvzvBtbSvgux4erkyS1e2hG3ZLJvmVkPyLv6UE7xYNZUtrj7tvj6zY6sdlW8ICU8fruN5OYuqKwBy_EYKNeBDiGY797vj2jB1972neTWX_73tEMN2cOFDfA7eRK6uyX0G9f0HsvnbxeH_dmioEyAyEd3wtm3ANAEtnZTthZ6rx9XXc36CFd70yTnbrwDSblvo2dLLgNxuhLDk678ekEAZfDcb34c0fF-VyOL8wgteQ7aZvygcErr1LgIWG7LnhGNhlu8GE3mOHBTN5vPXxbDiit4u1y_fiMPgov17ss3jXleXo5F6MWNnySY7GRh-tsB3redRU4ufEzBeYSlENMVpXAe7A0PCt3Hfpb2KzE3Rf8pt--F3le6W5uKdVp3zVrim3CMMOt7K2zDzRf9_jYjMDn5VS_-x4HLlfbcJfJA2LhvRGGR4L0ewrOHaO33ofID4kdHXtb8_cly22ih-yITJVbU5CxFwJ9DwxIWZ4RK1o2wfpRQjx63tXRud5hy9h3nsnQGPfx3OoYfLqsXzXQyPL15HkveA749HU53LxriJnviHfvcdGuj7h3nq30CKBaL-iUwXAcjGZpw9cx2s3sWgOIMCYzmF4ACR1vGppvgHmiRtj9WoAlxtgf8e3TOw62ul3JYi71bhSqMNWkHF-sG3u7e37L9x6dfSum0nO-Nzez4-zqoihq50pN6Vzfqp5ZHyybbG87dY8p85A7G5_n58vOBVoquqRyPDljbq7UvMvjvlyNvz1tPZ02BrhuPfPAXOiOd073Am-5tXDH3JVTJJcnVJPAfUvyLcudcbNz90IITaSymKVhzkPB94IyuWC5LOFXvsUbUb33JO26VrFQ2sftqu_N3Xtv9Z5BxjtjN-92ie6hO1qC5wraTDKOBeLzB6_vNkrvTNPDcSdHvXl5sBUctsbz-FhAM-AaqnKWBd3g8rRoOt9FO77FDI7xMAFIZBCfQh0QdQfmFW75Hmeb690LDCIJ5n3DmUbmeLQPqQu-OiFWXnUXoNmDqbCP41MbEdg3WsmnP4eqcNb0oOFECnndTITOYDn2ljcZyW-M914aHJKiEBW8M8Mq-gyUFka_7tPGgmm9oYi7mfHRmTcjbaf67vf4ViKE406q1q8766vL0HfHKiN038bPnYSOKlw5YTfK7H4L6lX7YBZ-3Cl8YwZhOTk_HlR8fO6-5nlo2UzfzCMxzzrelbB3lo3H8ol3Lm8Qv7eHDgyf-aaLLDnyB7PtzvVktl8KKJCanKgIby-nN9h7mNFttbBuGGicp-V7Z0at6azUlqN32h3yFCQ9M4CBeVy_va0dgxvPzBh8ke22uZSbkc7SDh9UwqHOiL6BlaGuXC02pemXzpt5Pj1Z0Usurs4jCfekDX2WuOMw-DBiDoe9G_jaLU2LNwPPs1fOzMs4oYu3P73ImXdjEkBGqVs-8awluUrbG8EKk7fd6LrlMkTw3mvgUK9l3sPO09SrsMlMrxLysmZpZGF2uUifid0CKt6-2Yt8IkzgXZKOKAqLNTXylNmtKbWCcRsernFc3z6KenzcWO_53nvk-HjYKjzr9M5DRNStVKL4cMiOsBt1dN7WsxkGSWMNNvukl41Fg8IN34Cabj1Ljyu5rrIa6fF5n-6pT4cKbEdVz2UfZlk47CJ-myCE9XabZAcRzT1MtI1s51nnB-yJ1Ny-7EPyEHLu3UPejIDxHXfIYUM58XtqtcMn3s4edzOR1HT8xsc7QWHr7RPltnyu2kN2Y6N6Wr7dbE-eVnnrVgdRpublMq7lKHLOL7tvEZztQmd0esJ0LWR8fK2mTh8sujn9llq8HBneGdwSXKJiLFS3KHrOO2365DWZ6b3rCyVLPLjwUk0ldrU36JyIV09Rz4EK3u6dF-DD7npm7KBC2r5nYfCwcmLR5WZIOuy5m_f8kPLu7akPe8M7do2qJ4hSvtfL07y7w0abSXtbVFdgehQFPmrBrfdsxVNT79st4BO3KTYeA74shYl2pzeFmvd81OKCBdSJxEE27303e33AWlzJtwCjVXDkW8nvEJJgi50FOZOeg7xdhdFL2ZV7ODHu4bxim72lnI1m4MfXjecuh-BlUyFCmde9l7g9-Uizp2ANSrV2uc6bnXFyhHjxIrbD3c5rMHxvujV44tsyj2z5hhuuLTXJ4k3BJYLW4dYh69tsJQf7WsHbQAYXIYQ3S33-8kWbl8nM25fS7Ro3oh6Ue4PNaKzyllPXYaoqx_LZRx9k49f2Pb_ze360GPTtGyLFXp1QfZPIcl4ChaF9m4P8nrHz8_vSKBi88at1Fd_bd9AMr_BFS6wXnOOr2u5il3mgXHl9Wp3vbuHX29vb_IinOXFeYVw8XfYg2C9biwI8ydedeyctnkXIkqluMPFYbeaOu3M78e6Ob6JVy3VN6MnMa3w8v6c4oqHLfO-9Yl-e9rmxe02oHZxFAO_IOAfzeee46GJzdHyDw2OdprEOyEYvDaF5fmNEhGacm9sBxQymNzOJbzKvo8rZE9kUj2lH7xlU8t57sW99730cg1xp0sd9pGrsbjYttifQdMoDh69H7t1mFuu3geS9AZDGg3u3BFvyxIe5GfgpPopq-lYYkRmRpZ3nHMNBSNxsRwduRndWLpHflzp3erOXKKs-PM6bzhgBA8ya2ITwU55vsfveeuzbY4bvzWHF98eLp-NlJZ74hqlH8cnNW5D38DzNFr6WGd8ilDyYE-6h-9Bo-zQbTWcmuWIo9HaKZWjRNYoBoTqDc9ZWpDtzMx1z9zz5o_z0eHM6aKwevMCdjY723T5k9IgM3rGWX8N5Ilq8DRQi8zxF50o9ZrroLOC2APHpuInKLvZh7727lm-FeLNHOuK2fmvqzb47P7y5ecuGSc51otel4nsHZYUkuy8-DzB70hrpzANrUx5T610beyr09IYLT613k3my-AK8R1_nyWLbAagGthyP1Rs8570BMiJfkAqNgpubSmmWzNsYL8vElqD3iELeEBx1MG8DeR2vnzNRSc2dMrdALFMfZyOdIh0v-97x-czrWqV1OTw57K0mjjMeWN820UORvN5biumem3SKp1JGRaYb2fequATrLmdnCL3VIu8xZcrDYwK99R_mfe9NvDHuYeQYm2Y49WA8DTScgx37OnuNjzp2z3TE6eR8v7xvZym-jmavhF1rlg9B9WIW-55VfGtBxzXb-AaSLkHREdEe1fNBfImcnKjunUjd3ts7Bc8eAd8bWfb27PJZnil-wZdg7DJqikRNCivGqmzv1lrsBLbf3LZKRtyv3rxshpSm6YiAtBDDyvVeJnhE-qSHMRZVLTZ3e0j2HqZwLFYcOfeiV747janrKLe3DSYSUDvnTAGb2eLIZ-iztp7eM-PA9uSgmZm30X7fapLxLNkyFwUHpqdHb99YNkYXMxdjI22-3npwO0nUPofZdyswbbDAklFRKm-4j43vsg5zknxBTd0dD297-otQ}]
\label{ex:moving_intruder}
\begin{lstlisting}
import math
input lat: Float
input lon: Float
input intruder_lat: Float
input intruder_lon: Float

output distance @((intruder_lat && intruder_lon) || (lat && lon))@ :=  sqrt((intruder_lat.hold(or: 0.0) - lat.hold(or: 0.0))**2.0 + (intruder_lon.hold(or: 0.0) - lon.hold(or: 0.0))**2.0)
output closer ?@((intruder_lat && intruder_lon) || (lat && lon))@? :=
	distance.offset(by: -1).defaults(to: distance) >= distance 

trigger @1Hz@ closer.aggregate(over_exactly: 5s, using: forall).defaults(to: false) && distance.hold(or: 1.0) < 0.1 "Too close to the intruder"
\end{lstlisting}
\end{example}
While the \stream{distance} stream mathematically performs the same computation, a \lstinline{hold()} lookup is used to avoid a direct dependency.
This lookup refers to the most current available value of the accessed stream and does not require that the accessed value is computed at the same time.
However, there might not be such a value when the accessing stream is evaluated, so a default value must be supplied similar to the offset operator.
As the timing of the \stream{distance} stream is decoupled from any input stream, it has to be explicitly specified when it should produce new values.
Syntactically, this is specified through a positive boolean expression over input streams following the \lstinline{@} after a stream's name.
This expression is called the \emph{activation condition} of the stream and symbolically describes the events at which the stream is evaluated.
In the example, the \stream{distance} stream is evaluated whenever the intruders \emph{or} the drone's position changes.
For the \stream{closer} stream the activation condition is the same as for the \stream{distance}, but is automatically inferred because of the synchronous access.

Moreover, the trigger in the specification has changed.
It is now periodic, a concept which will be introduced in the subsequent section.

\subsection{Periodic Streams}
In reality, it is often required that a monitor not only reacts to system actions but can also proactively produce verdicts about the system's health.
Otherwise, the monitor could not detect a frozen system because it would freeze as well.
In \rtlola, proactive monitoring is achieved through streams evaluated at a fixed frequency called periodic streams.
A periodic stream can be specified by giving a frequency or period annotation like \lstinline{1Hz} or \lstinline{1s} after the \lstinline{@} keyword.
These frequencies are independent of input streams and to access input streams from periodic output streams we can either use \lstinline|hold|-accesses or sliding windows.
In our example, we use a sliding window in the trigger stream to make the specification more robust against GPS fluctuations.

\begin{figure}[t]
	\begin{center}
        \def\distance{0.8}
        \def\offset{0.2}
        \begin{tikzpicture}[circ/.style={draw,circle,fill=white,inner sep=0.5mm}]
            \node[anchor=east] (a) {\texttt{closer}};
            \node[below=14mm of a.east,anchor=east] (x) {\texttt{trigger}};
            \draw (a.east) -- ++(5.7cm,0) coordinate (right);
            \draw (x.east) -- ++(5.7cm,0) coordinate (right);
            \foreach \i/\v/\a in {0.7/F/1,2.7/T/2,4/T/3,5.4/T/4,6.5/F/5}{
                \node[circ] (a\a) at ($(a.east)+(\i*\distance-\offset,0)$) {\scriptsize\v};
            }
            \foreach \i in {1,2,3,4}{
                \node[circ,fill=lightgray] (x\i) at ($(x.east)+(\i*\distance-\offset,0)$) {\scriptsize F};
            }
            \foreach \i/\v in {5/F,6/T,7/F}{
                \node[circ] (x\i) at ($(x.east)+(\i*\distance-\offset,0)$) {\scriptsize\v};
            }
            \coordinate (LL) at ([yshift=-3.5mm]x.east);
            \draw ([yshift=2mm]a.east) -- (LL);
            \draw[->] (LL) -- ++ (5.8cm,0) node[right] {$t[s]$};
            \foreach \i in {1,2,3,4,5,6,7}{
                \node at ($(LL)+(\i*\distance-\offset,-2mm)$) {\scriptsize$\i$};
            }
            \begin{scope}[on background layer]
            \clip (a.east) rectangle ($(x7 |- LL) + (1mm,0)$);
            \filldraw[red,fill=red!50!white,fill opacity=0.2] (x5.center) -- ($(x5.center |- a) - (5*\distance,0)$) -- (x5 |- a) -- cycle;
            \filldraw[green,fill=green!50!white,fill opacity=0.2] (x6.center) -- ($(x6.center |- a) - (5*\distance,0)$) -- (x6 |- a) -- cycle;
            \filldraw[blue,fill=blue!50!white,fill opacity=0.2] (x7.center) -- ($(x7.center |- a) - (5*\distance,0)$) -- (x7 |- a) -- cycle;
            \end{scope}
        \end{tikzpicture}
	\end{center}
	\caption{The functioning of sliding windows exemplified based on \Cref{ex:moving_intruder}}
	\label{fig:sliding_window}
\end{figure}
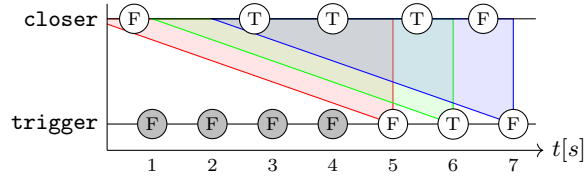

Sliding window aggregations aggregate over every value in a given time frame of a stream using an aggregation function.
More concretely, the window in our example evaluates only to true if every \lstinline|closer| value in the last 5 seconds is true.
The functionality of this sliding window is visualized in \Cref{fig:sliding_window}.
In our specification, the trigger stream is evaluated with a fixed frequency of one second whereas the \lstinline|closer| stream depends on the inputs.
For the first four seconds, the window does not aggregate any values given that  the whole duration is not available at the start of the monitoring.
In this case, the window evaluates to the default value \lstinline{false} instead.
Afterwards, the window aggregates over different numbers of closer values as illustrated by the different colors.

\rtlola supports a set of aggregation functions as \lstinline{exists, avg, sum, count, min} or \lstinline{max}.

\begin{expertSection}{\rtlola's Type System}
Similar to many programming languages, \rtlola has a type system that ensures only valid specifications can be executed.
The type-system of \rtlola is twofold:
The value type of a stream describes how the memory should be interpreted, i.e., as a signed or unsigned integer, a floating point number, or a string.
The underlying value type system ensures that only compatible values are used in an operation, such that, for example, no string and number are added.
The pacing type of a stream describes its timing behavior, i.e., it determines the points in time when the stream is evaluated.
This pacing type system ensures that every direct access, also called synchronous access, is valid.
Consider the following example specification with an invalid synchronous access:
\begin{example}[\spec{Invalid Synchronous Access}{https://rtlola.cispa.de/playground/?spec=G0AA-I3EOBbxgTBQUIKaHJfQCUctsn9uXvnWo-MxdmUYd1nMLS1AAjOso9GiGDaRfeSPFjUST-cA&trace_name=c2ltcGxlX3RyYWNl&trace=iwuAYSxiLHRpbWUKZmFsc2UsZmFsc2UsMC4wAw==}]
\label{ex:wrong_pt}

\begin{minipage}{0.5\linewidth}
    \begin{center}
    \begin{tabular}{c}
    \begin{lstlisting}
        input a: Bool
        input b: Bool
        output x @a@ := a
        output y @b@ := x
    \end{lstlisting}
    \end{tabular}
    \end{center}
    \end{minipage}
    \begin{minipage}{0.5\linewidth}
        \centering
        \def\distance{0.55}
        \scalebox{0.9}{
        \begin{tikzpicture}[circ/.style={draw,circle,fill=white,inner sep=0.8mm},d/.style={dashed,dash pattern=on 2pt off 2pt}]
            \node[anchor=east] (a) {\texttt{a}};
            \node[below=4mm of a.east,anchor=east] (b) {\texttt{b}};
            \node[below=4mm of b.east,anchor=east] (x) {\texttt{x}};
            \node[below=7mm of x.east,anchor=east] (y) {\texttt{y}};
            \draw (a.east) -- ++(3.1cm,0);
            \draw (b.east) -- ++(3.1cm,0);
            \draw (x.east) -- ++(3.1cm,0);
            \draw (y.east) -- ++(3.1cm,0);
            \foreach \i in {1,2,4}{
                \node[circ] (a\i) at ($(a.east)+(\i*\distance,0)$) {};
            };
            \foreach \i in {1,2,3,5}{
                \node[circ] (b\i) at ($(b.east)+(\i*\distance,0)$) {};
            };
            \foreach \i in {1,2,4}{
                \node[circ] (x\i) at ($(x.east)+(\i*\distance,0)$) {};
            };
            \foreach \i in {1,2}{
                \node[circ] (y\i) at ($(y.east)+(\i*\distance,0)$) {};
            };
            \node[circ,d] (y3) at ($(y.east)+(3*\distance,0)$) {};
            \node[circ,d] (y5) at ($(y.east)+(5*\distance,0)$) {};

            \node[circ,fill=red!50,minimum width=2.5mm,inner sep=0] (x5) at ($(x.east)+(5*\distance,0)$) {\tiny ?};
            \node[circ,fill=red!50,minimum width=2.5mm,inner sep=0] (x3) at ($(x.east)+(3*\distance,0)$) {\tiny ?};
            \draw[->,shorten >=0.5mm] (y3) -- (x3);
            \draw[->,shorten >=0.5mm] (y5) -- (x5);
            \draw[->,shorten >=0.5mm,d,darkgray] (y3) -- (x2);
            \draw[->,shorten >=0.5mm,d,darkgray] (y5) -- (x4);
            \coordinate (LL) at ([yshift=-3.5mm]y.east);
            \draw ([yshift=2mm]a.east) -- (LL);
            \draw[->] (LL) -- ++ (3.3cm,0) node[right] {$t[s]$};
            \foreach \i/\a in {1/0.7,2/1.0,3/1.2,4/1.4,5/1.8}{
                \node at ($(LL)+(\i*\distance,-2mm)$) {\scriptsize$\a$};
            }
        \end{tikzpicture}}
    \end{minipage}
    
\end{example}
The specification on the left has two output streams: The stream \stream{x}, evaluated whenever the input \stream{a} receives a new value, with that same value of \stream{a}.
The stream \stream{y} is evaluated whenever the input \stream{b} receives a new value and takes the value of \stream{x}.
The diagram on the right exemplifies the timing behavior of the streams.
At time $1.2$ and $1.8$, only stream \stream{b} receives a new value.
Consequently, the stream \stream{x} is not computed, and with that stream \stream{y} cannot access the value of \stream{x} at these points in time.
These timing errors can lead to invalid memory accesses at runtime, so the type checker rejects the specification.
To fix the specification, one could use a \lstinline{hold} access from \stream{y} to \stream{x} similar to \Cref{ex:moving_intruder}, resulting in the greyed-out arrows in the diagram on the right.

\paragraph{The Pacing Type System}
Intuitively, the pacing type system declares a synchronous access valid if the accessed stream is evaluated at least at the same points in time as the accessing stream.
For non-periodic, also called event-based streams, this property can be checked by asserting logical implication between the activation condition of the accessing stream and the accessed stream.
In \Cref{ex:wrong_pt}, it is easy to see that $a \rightarrow b$ does not always hold, but, for example, $a \land b \rightarrow a$ does.
A similar condition holds for periodic streams:
A synchronous access between periodic streams is valid if the accessing stream runs at a slower pace than the accessed stream.
Concretely, the period of the accessed stream has to be a multiple of the period of the accessing stream.
In addition, synchronous accesses between periodic and event-based streams are never valid.

\end{expertSection}

\section{Stream Lifecycle}
\label{sec:lifecycle}
In this section, we optimize the specification from \Cref{ex:moving_intruder} by guarding computationally heavy operations with cheap-to-evaluate predicates.
Concretely, we only compute the \stream{distance} and \stream{closer} streams and the trigger when the intruder is in range.
These optimizations are enabled by lifting the assumption that a stream exists from the start to the end of the monitor.
Instead, we allow for dynamic stream creation, i.e., a stream can be created and removed from the monitor at runtime.
In the following, we refer to the creation of streams as their spawn behavior and their deletion as their close behavior.
In between its spawn and close action, a stream is evaluated as defined by its stream expression.

Syntactically, the three steps of a stream's lifecycle are specified by three sub-clauses in the stream's definition:
\begin{lstlisting}
	output $o$
		spawn (*\color{greentypes}@$p_s$*) when $c_s$
		eval (*\color{greentypes}@$p_e$*) when $c_e$ with $e$
		close (*\color{greentypes}@$p_c$*) when $c_c$
\end{lstlisting}
Each of these clauses can feature a pacing and a boolean stream expression preceded by the \lstinline{when} keyword.
The pacing of a clause \emph{statically} determines when the clause could be evaluated, i.e., whenever event \lstinline{a&b} occurs or at a frequency of \lstinline{1Hz}.
The stream expression preceded by \lstinline{when} is evaluated at the time points described by the pacing.
It constrains these time points dynamically based on runtime values, i.e., the action corresponding to the clause is only taken \emph{when} this expression evaluates to true.
The evaluation clause features a stream expression defining \emph{how} the stream's value is computed after the \lstinline{with} keyword.

As for the previous sections, pacings can be omitted and inferred by the type checker for brevity.
The \lstinline{when} expression of a clause can also be omitted, representing the constant \lstinline{true}.
If a stream's spawn or close clause is omitted, the stream exists from the start or until the end of the monitor, respectively.
If only an \lstinline|eval with| clause exists, we can use the short-hand notation \lstinline|:=| as used in the previous sections.

Consider the following extension of \Cref{ex:moving_intruder} in which every stream in the specification is now composed of three clauses:
a spawn clause, an eval clause, and a close clause.
\begin{example}[\spec{An Out-of-Range Intruder}{https://rtlola.cispa.de/playground/?spec=G-EDAIzDOIY8uChtivbpaG6ZL-nAJyIJKWnmzCRA_udmAF4h_cYYIFkBrWFxAY5puc57frbZXkH0qI_T1NPZwN-IAJGDMPNI5K-h-E_DHCeDVweT_q7bYBpFaP_57GkuAWFFrVeuIQAYu2D9g_vS2IfHQw-PccMXDbv7wmeL-BoEGiHOnWrlCIz9ML0DdlDj7nbQWZrFgAL6OYbjPM0CAbIo7R-1_RMm4GWWc0MC68BgnII6Ocm0FjGMCInWhxs8pSEm_g6NSLjrFIfRcQ7H19TqjD5mW4Sa7SUYar88l2P71xkheldDIE1VnZMn-2m83pVbDbNYY0U0TVKGNpNHoIhGmH44LgkGJktzHhrfwzAioHm-iV7x6OiSi4dpOwR1HSvKEQIs8_I7fhTEyZREUxJfwNAQ&trace_name=c2ltcGxlX3RyYWNl&trace=G5gtEZW6EkBngm2jceuWNYTa2IfhJVvYm9MW5d_Yt5fNUIDSJnOTo9MsX2V1_K9Vzna32kRFclwyclw8QqFxWBQVx3ELR1EehDxu43CuZ957vzfpnUQo1jj8n1ZTLeGp5VN7Nd31BzY5QVQXg9NExKuat379Gvb5x6__aYOPU36nX1-_fT35ob-c5-fJPzD8D43um4adpN6cf7pvjJspdpBH_YbfMa5z2fHLb7-RT6b67vF-_1--f99RfT8xvVfY655u6ew11JPCxYO5UWYRKI_HGWLfForZuyNxBgR69T5UJ52eaLeQgbSnUQHPTvgUPgcpE8PRPTPP5ud7pN75rJ7fLKkusxf6oSKCITs2OJSjvBJVRqgU16FWYOW9ARsolkXzoYHN6VPRFe_lBaBppM_0MDnx8c5rB306QTOcfVi-nUKHaiFTb4kcaktZ4bgG8W4zg81auTO30i2zFTyTUNKd2wdSuV5v1s9Bd5p5zt3Az8ri-Z4Yl8RjvjWOz0rIzHuT17kbm4u1HtRPJ1NhBs6ym1rvnHvPN-Vjylpe02RFJy1FfYkOuXMNj-9mzmZRQ294COQMcPvWVMpTdC6m4hU6Q769jLB5YwpYHuzlQZ3XqLs8uIvqPd0dqvzu3no81DvhLniP6PNj5QHlQhp4OWH3wD5vOT59WShzbcolcw9HkKhA7_C-WMQKJN_rVQpS5rGGuru4dHoqK2UyidW3fAte8SBvIEHeEkP7CSZfX9mLLwVXjZYoJq1kb0VOlaGsMNYLDKJ3ahgd5ZoLGC35bnF319OEmQyGHmMCgGPEkyw_6evobhcVeBFx0W7vPSh_3_zsNxS5FiCH3L1TmBV824wzxq_Va_vejPMey-W-XN7E73NJzL0WdTv-HFEQkXY7LpqZtx5BKLd370nscH41Pqh6w-UC3qTWy9OSmh4rKzDlnhAzh9m3hHl5xJnTTfLHpxv3Na4gu0X7Ng83RfM67ZaUYBd-M3hdulhX-T2Ow7UCl-4-XZep5OZXnPy41SjsM87oKzK38b0mnc5pcLPRpup689bF3B70gyn2vQKbDUmaQW6UUrPeszUSnMfo5o6DctVUodwKiWk7w17eJAAJr8IeckNAPZIPtzzvvawHWNYWgJ7sM5k6IfY9vNu5TSr0mddbkzuoFteKukPwsH18M0euZSs-Ll_kOD3veqtX1cKEpXdFmruUPCgHbgDzFWCBPUt-XnyzWr82cSHwQpxmdTMzt0JJlS8L4RGeJ-NQdutRbe76Qt76VhvToN_jjPfKe9u7-iQ2g8cFezZhwB48MAqrSVf2zopHgCD0tSM21-REUWJzz3APo59mVjsOtgiHqAeKzxzZEN8DxGuHEZ3JlREzhyUW91jkCTECr3DgzWaooW3hPULz3vrohisF2Pee9xoKPmpxYxifnt9U-150MU6n097dWFmzT3srv6nDeS1X3UWy6mEh7Rn3DchT8TBWCu8gKhMuc85U4dww93S_eUv5VrdvZIZuaXdfbk2ImCRvkF08zdu89OZpl3HcatLsrJ53farF43tWFAmcOIegbRn2uqPBBme0jsFp3gte7jD3IkILJ8pajVjqYXj1aQdHLlQ2j8-lfvZUwzb1DUo_PUqEfvM9siFfnmyOku_O97ASNlOVWG_gPq2e--zEaHzRXEcz0sjlVYtyptQM4D3cIexdd4p7e4Ohxc08rbR8eNfNzeKeRV7VWz2k2sdVMdc3z2_BtTy4AUZwX0_zkAx28ysXZ6XaiuTBcnDp2wbf4_n5oXcNftZM0FtKM40thhhd5ax4d_fOpKqVoI7vdRGXY5wZBB9oBmi6897QPYPc2XvzvBtbSvgux4erkyS1e2hG3ZLJvmVkPyLv6UE7xYNZUtrj7tvj6zY6sdlW8ICU8fruN5OYuqKwBy_EYKNeBDiGY797vj2jB1972neTWX_73tEMN2cOFDfA7eRK6uyX0G9f0HsvnbxeH_dmioEyAyEd3wtm3ANAEtnZTthZ6rx9XXc36CFd70yTnbrwDSblvo2dLLgNxuhLDk678ekEAZfDcb34c0fF-VyOL8wgteQ7aZvygcErr1LgIWG7LnhGNhlu8GE3mOHBTN5vPXxbDiit4u1y_fiMPgov17ss3jXleXo5F6MWNnySY7GRh-tsB3redRU4ufEzBeYSlENMVpXAe7A0PCt3Hfpb2KzE3Rf8pt--F3le6W5uKdVp3zVrim3CMMOt7K2zDzRf9_jYjMDn5VS_-x4HLlfbcJfJA2LhvRGGR4L0ewrOHaO33ofID4kdHXtb8_cly22ih-yITJVbU5CxFwJ9DwxIWZ4RK1o2wfpRQjx63tXRud5hy9h3nsnQGPfx3OoYfLqsXzXQyPL15HkveA749HU53LxriJnviHfvcdGuj7h3nq30CKBaL-iUwXAcjGZpw9cx2s3sWgOIMCYzmF4ACR1vGppvgHmiRtj9WoAlxtgf8e3TOw62ul3JYi71bhSqMNWkHF-sG3u7e37L9x6dfSum0nO-Nzez4-zqoihq50pN6Vzfqp5ZHyybbG87dY8p85A7G5_n58vOBVoquqRyPDljbq7UvMvjvlyNvz1tPZ02BrhuPfPAXOiOd073Am-5tXDH3JVTJJcnVJPAfUvyLcudcbNz90IITaSymKVhzkPB94IyuWC5LOFXvsUbUb33JO26VrFQ2sftqu_N3Xtv9Z5BxjtjN-92ie6hO1qC5wraTDKOBeLzB6_vNkrvTNPDcSdHvXl5sBUctsbz-FhAM-AaqnKWBd3g8rRoOt9FO77FDI7xMAFIZBCfQh0QdQfmFW75Hmeb690LDCIJ5n3DmUbmeLQPqQu-OiFWXnUXoNmDqbCP41MbEdg3WsmnP4eqcNb0oOFECnndTITOYDn2ljcZyW-M914aHJKiEBW8M8Mq-gyUFka_7tPGgmm9oYi7mfHRmTcjbaf67vf4ViKE406q1q8766vL0HfHKiN038bPnYSOKlw5YTfK7H4L6lX7YBZ-3Cl8YwZhOTk_HlR8fO6-5nlo2UzfzCMxzzrelbB3lo3H8ol3Lm8Qv7eHDgyf-aaLLDnyB7PtzvVktl8KKJCanKgIby-nN9h7mNFttbBuGGicp-V7Z0at6azUlqN32h3yFCQ9M4CBeVy_va0dgxvPzBh8ke22uZSbkc7SDh9UwqHOiL6BlaGuXC02pemXzpt5Pj1Z0Usurs4jCfekDX2WuOMw-DBiDoe9G_jaLU2LNwPPs1fOzMs4oYu3P73ImXdjEkBGqVs-8awluUrbG8EKk7fd6LrlMkTw3mvgUK9l3sPO09SrsMlMrxLysmZpZGF2uUifid0CKt6-2Yt8IkzgXZKOKAqLNTXylNmtKbWCcRsernFc3z6KenzcWO_53nvk-HjYKjzr9M5DRNStVKL4cMiOsBt1dN7WsxkGSWMNNvukl41Fg8IN34Cabj1Ljyu5rrIa6fF5n-6pT4cKbEdVz2UfZlk47CJ-myCE9XabZAcRzT1MtI1s51nnB-yJ1Ny-7EPyEHLu3UPejIDxHXfIYUM58XtqtcMn3s4edzOR1HT8xsc7QWHr7RPltnyu2kN2Y6N6Wr7dbE-eVnnrVgdRpublMq7lKHLOL7tvEZztQmd0esJ0LWR8fK2mTh8sujn9llq8HBneGdwSXKJiLFS3KHrOO2365DWZ6b3rCyVLPLjwUk0ldrU36JyIV09Rz4EK3u6dF-DD7npm7KBC2r5nYfCwcmLR5WZIOuy5m_f8kPLu7akPe8M7do2qJ4hSvtfL07y7w0abSXtbVFdgehQFPmrBrfdsxVNT79st4BO3KTYeA74shYl2pzeFmvd81OKCBdSJxEE27303e33AWlzJtwCjVXDkW8nvEJJgi50FOZOeg7xdhdFL2ZV7ODHu4bxim72lnI1m4MfXjecuh-BlUyFCmde9l7g9-Uizp2ANSrV2uc6bnXFyhHjxIrbD3c5rMHxvujV44tsyj2z5hhuuLTXJ4k3BJYLW4dYh69tsJQf7WsHbQAYXIYQ3S33-8kWbl8nM25fS7Ro3oh6Ue4PNaKzyllPXYaoqx_LZRx9k49f2Pb_ze360GPTtGyLFXp1QfZPIcl4ChaF9m4P8nrHz8_vSKBi88at1Fd_bd9AMr_BFS6wXnOOr2u5il3mgXHl9Wp3vbuHX29vb_IinOXFeYVw8XfYg2C9biwI8ydedeyctnkXIkqluMPFYbeaOu3M78e6Ob6JVy3VN6MnMa3w8v6c4oqHLfO-9Yl-e9rmxe02oHZxFAO_IOAfzeee46GJzdHyDw2OdprEOyEYvDaF5fmNEhGacm9sBxQymNzOJbzKvo8rZE9kUj2lH7xlU8t57sW99730cg1xp0sd9pGrsbjYttifQdMoDh69H7t1mFuu3geS9AZDGg3u3BFvyxIe5GfgpPopq-lYYkRmRpZ3nHMNBSNxsRwduRndWLpHflzp3erOXKKs-PM6bzhgBA8ya2ITwU55vsfveeuzbY4bvzWHF98eLp-NlJZ74hqlH8cnNW5D38DzNFr6WGd8ilDyYE-6h-9Bo-zQbTWcmuWIo9HaKZWjRNYoBoTqDc9ZWpDtzMx1z9zz5o_z0eHM6aKwevMCdjY723T5k9IgM3rGWX8N5Ilq8DRQi8zxF50o9ZrroLOC2APHpuInKLvZh7727lm-FeLNHOuK2fmvqzb47P7y5ecuGSc51otel4nsHZYUkuy8-DzB70hrpzANrUx5T610beyr09IYLT613k3my-AK8R1_nyWLbAagGthyP1Rs8570BMiJfkAqNgpubSmmWzNsYL8vElqD3iELeEBx1MG8DeR2vnzNRSc2dMrdALFMfZyOdIh0v-97x-czrWqV1OTw57K0mjjMeWN820UORvN5biumem3SKp1JGRaYb2fequATrLmdnCL3VIu8xZcrDYwK99R_mfe9NvDHuYeQYm2Y49WA8DTScgx37OnuNjzp2z3TE6eR8v7xvZym-jmavhF1rlg9B9WIW-55VfGtBxzXb-AaSLkHREdEe1fNBfImcnKjunUjd3ts7Bc8eAd8bWfb27PJZnil-wZdg7DJqikRNCivGqmzv1lrsBLbf3LZKRtyv3rxshpSm6YiAtBDDyvVeJnhE-qSHMRZVLTZ3e0j2HqZwLFYcOfeiV747janrKLe3DSYSUDvnTAGb2eLIZ-iztp7eM-PA9uSgmZm30X7fapLxLNkyFwUHpqdHb99YNkYXMxdjI22-3npwO0nUPofZdyswbbDAklFRKm-4j43vsg5zknxBTd0dD297-otQ}]
\label{ex:out-of-range}
\begin{lstlisting}
import math
input intruder_lat: Float
input intruder_lon: Float
input lat: Float
input lon: Float

output distance 
    spawn @(intruder_lat && intruder_lon)@
    eval @((intruder_lat && intruder_lon) || (lat && lon))@
    	with  sqrt((intruder_lat.hold(or: 0.0) - lat.hold(or: 0.0))**2.0 + (intruder_lon.hold(or: 0.0) - lon.hold(or: 0.0))**2.0)
    close @true@ when stale.hold(or: false)
output closer
    spawn @(intruder_lat && intruder_lon)@
    eval with distance.offset(by: -1).defaults(to: distance) >= distance
    close @true@ when stale.hold(or: false)
output stale
    spawn @(intruder_lat && intruder_lon)@
    eval @10s@ with intruder_lat.aggregate(over: 10s, using: count) = 0
    close @true@ when stale.hold(or: false)

trigger
    spawn @((intruder_lat && intruder_lon) || (lat && lon))@
    	when distance.hold(or: 1.0) < 0.1
    eval @1Hz@ when closer.aggregate(over_exactly: 5s, using: forall).defaults(to: false) with "Intruder detected"
    close @true@ when stale.hold(or: false) 
\end{lstlisting}
\end{example}
Another notable change is the added \stream{stale} stream.
It defines when the intruder is considered out of range by checking for any GPS coordinate updates in the last 10 seconds.
The \lstinline{spawn} clause of the stream defines that this condition is only monitored once a GPS location is received from the intruder.
The omitted \lstinline{when} expression in the spawn clause is equivalent to a \lstinline{when true} definition.
The \lstinline{close} clause of the stream defines that the condition should no longer be monitored as soon as it becomes true for the first time.
However, the stream is spawned again if a new intruder GPS location reactivates the spawn condition of the stream.

The \stream{distance} and \stream{closer} streams inherit the same \lstinline{spawn} and \lstinline{close} clauses as the \stream{stale} stream, further excluding redundant computations.
The trigger also inherits the same close condition as the other two streams.
Yet, its spawn condition differs slightly.
The \lstinline{distance.hold(or: 1.0) < 0.1} expression is moved from the main trigger condition to its spawn condition, adapting the spawn activation condition accordingly.
That way, the computationally heavy sliding window aggregation is only performed once the intruder is close enough.

Besides optimizing specifications, dynamic stream creations increase the expressiveness of the \rtlola specification language as shown in the next subsection.

\subsection{Deadline Watchdogs}
Deadline watchdogs are common specification requirements in CPS and can be expressed in natural language as follows: "$t$ seconds after event $e$ a condition $c$ must hold."
Such requirements can be represented in \rtlola using dynamically created streams:
\begin{example}[A Deadline Watchdog]
	\label{ex:watchdog}
	\begin{lstlisting}
		output timer
			spawn when $e$
			eval $\color{greentypes}\texttt{@}t\texttt{s}$ with true
			close when timer
			
		trigger
			spawn when $e$
			eval $\color{greentypes}\texttt{@}t\texttt{s}$ when !$c$ with "The deadline was missed"
			close when timer
	\end{lstlisting}
\end{example}
For the watchdog, we define a new stream \lstinline!timer! that is spawned with the start of the watchdog, i.e. the event $e$ spawns the watchdog.
This is the starting point of the annotated frequency, so this stream is evaluated for the first time $t$ seconds after $e$.
Since this value is immediately true, we also close the stream after these $t$ seconds.
The trigger stream has the same \lstinline|spawn| and \lstinline|close| condition and is evaluated with the same frequency.
Here, we check if the condition $c$ is satisfied and use the \lstinline|timer| stream as a helper function to immediately close the trigger stream after the first evaluation.

		

\begin{expertSection}{Semantic Types}
The introduced \lstinline{when}-conditions further refine the timing of a stream and add another point of failure for synchronous accesses.
This problem is solved with another type system.

\paragraph{Semantic Types and Event-based Streams}
The following example illustrates a possible point of failure using synchronous accesses and \lstinline|when|-conditions:
\begin{example}[\spec{Invalid Semantic Types}{https://rtlola.cispa.de/playground/?spec=G1kAqAQ-f8c3oQ--0aCXJnggUbCcciABU0Helkh8sMMO2Gkttg_JuLg594WChiGkinSbyWZR12FL7wfJAQ==&trace_name=c2ltcGxlX3RyYWNl&trace=iwWAYSx0aW1lCjEsMC4wAw==}]
\label{ex:wrong_sem_type}

\begin{minipage}{0.5\linewidth}
\begin{lstlisting}
	input a: Int
	output x 
		eval @a@ when a > 4 with a
	output y
		eval @a@ when a > 3 with x
\end{lstlisting}
\end{minipage}
\begin{minipage}{0.45\linewidth}
\def\distance{0.7}
    \centering
    \scalebox{0.9}{
\begin{tikzpicture}[circ/.style={draw,circle,fill=white,inner sep=0.5mm}]
    \node[anchor=east] (a) {\texttt{a}};
    \node[below=6mm of a.east,anchor=east] (x) {\texttt{x}};
    \node[below=6mm of x.east,anchor=east] (y) {\texttt{y}};
    \draw (a.east) -- ++(3.1cm,0);
    \draw (x.east) -- ++(3.1cm,0);
    \draw (y.east) -- ++(3.1cm,0);
    \foreach \i/\v in {1/5,2/1,3/8,4/4}{
        \node[circ] (a\i) at ($(a.east)+(\i*\distance,0)$) {\scriptsize\v};
    };
    \foreach \i/\v in {1/5,3/8}{
        \node[circ] (x\i) at ($(x.east)+(\i*\distance,0)$) {\scriptsize\v};
		\draw[->,shorten >=0.5mm] (x\i) -- (a\i);
    };
    \foreach \i/\v in {1/5,3/8}{
        \node[circ] (y\i) at ($(y.east)+(\i*\distance,0)$) {\scriptsize\v};
		\draw[->,shorten >=0.5mm] (y\i) -- (x\i);
    };
    \node[circ,dashed] (y3) at ($(y.east)+(4*\distance,0)$) {\phantom{\scriptsize ?}};
    \node[circ,fill=red!50] (x3) at ($(x.east)+(4*\distance,0)$) {\scriptsize ?};
	\draw[->,shorten >=0.5mm] (y3) -- (x3);
    \coordinate (LL) at ([yshift=-3.5mm]y.east);
    \draw ([yshift=2mm]a.east) -- (LL);
    \draw[->] (LL) -- ++ (3.2cm,0) node[right] {$t[s]$};
    \foreach \i in {1,2,3,4}{
        \node at ($(LL)+(\i*\distance,-2mm)$) {\scriptsize$\i$};
    }
\end{tikzpicture}}
\end{minipage}
\end{example}
As specified by the \lstinline{when} expression of the \lstinline{eval} clause of \stream{x}, it only produces a value when the input \stream{a} is greater than 4.
This might not coincide with \lstinline|a > 3| as required for the evaluation of \stream{y}.
Therefore, similar to \Cref{ex:wrong_pt}, there are points in time when \stream{y} is evaluated, but \stream{x} is not.
To detect specifications with such errors, \rtlola uses another type system reasoning about the when conditions, called semantic types in that context.
Like the pacing type system, the semantic type system ensures the implication relation between semantic types holds.
In the above example, the type system would reason whether the implication $ a > 3 \rightarrow  a > 4$ is a tautology and consequently reject this specification.

Besides ensuring that the \emph{when} conditions of the eval clauses of dependent streams imply each other, the semantic type system also reasons about the lifecycle of dependent streams.
Concretely, for a synchronous access to succeed, the accessed stream must be alive at least as long as the accessing stream.

\paragraph{Semantic Types and Periodic Streams}
While the previous intuition holds for event-based streams, synchronous accesses between periodic streams imply stricter requirements.
Consider the following example:
\begin{example}[\spec{Shifted Periodicity}{https://rtlola.cispa.de/playground/?spec=G3gA4MTylu9L-qox0gcTKfFDeYYNPOAiQh9zIo0BNuCImk-WBWBZhZttgtD8JWANPTawi8yRgFZxCfXLB_WBjOpT2LsgeCDxAg==&trace_name=c2ltcGxlX3RyYWNl&trace=iwWAYSx0aW1lCjEsMC4wAw==}]

\begin{minipage}{0.5\linewidth}
	\begin{lstlisting}
	input a: Int
	output x
		spawn when a > 3
		eval @1Hz@ with a.hold(or: 0)
	output y
		spawn when a > 4
		eval @1Hz@ with x
	\end{lstlisting}
\end{minipage}
\begin{minipage}{0.45\linewidth}
\def\distance{0.9}
\def\offset{0.3}
    \centering
    \scalebox{0.9}{
\begin{tikzpicture}[circ/.style={draw,circle,fill=white,inner sep=0.5mm}]
    \node[anchor=east] (a) {\texttt{a}};
    \node[below=6mm of a.east,anchor=east] (x) {\texttt{x}};
    \node[below=6mm of x.east,anchor=east] (y) {\texttt{y}};
    \draw (a.east) -- ++(3.6cm,0) coordinate (right);
	\draw[Bracket-] ($(x.east)+(2*\distance-\offset,0)$) -- (right |- x);
	\draw[Bracket-] ($(y.east)+(2.5*\distance-\offset,0)$) -- (right |- y);
    \foreach \i/\v/\a in {1/1/1,2/4/2,2.5/5/p,3.5/2/q}{
        \node[circ] (a\a) at ($(a.east)+(\i*\distance-\offset,0)$) {\scriptsize\v};
    };
    \foreach \i/\v/\a in {3/5/p,4/2/q}{
        \node[circ] (x\i) at ($(x.east)+(\i*\distance-\offset,0)$) {\scriptsize\v};
		\draw[->,shorten >=0.5mm] (x\i) -- (a\a);
    };
    \foreach \a/\i in {b/3.5}{
        \node[circ,dashed] (y\a) at ($(y.east)+(\i*\distance-\offset,0)$) { \phantom{\scriptsize ?}};
        \node[circ,fill=red!50] (x\a) at ($(x.east)+(\i*\distance-\offset,0)$) {\scriptsize ?};
		\draw[->,shorten >=0.5mm] (y\a) -- (x\a);
    };
    \coordinate (LL) at ([yshift=-3.5mm]y.east);
    \draw ([yshift=2mm]a.east) -- (LL);
    \draw[->] (LL) -- ++ (3.7cm,0) node[right] {$t[s]$};
    \foreach \i in {1,1.5,2,2.5,3,3.5,4}{
        \node at ($(LL)+(\i*\distance-\offset,-2mm)$) {\scriptsize$\i$};
    }
\end{tikzpicture}}
\end{minipage}
\end{example}
Here the streams \stream{x} and \stream{y} have the same frequency and the spawn condition of \stream{y} implies the spawn condition of \stream{x}.
However, the synchronous access might fail if \stream{x} spawns before \stream{y} since the frequencies are now out of sync.
For example, the sequence of events depicted on the right of the example will lead to an invalid synchronous access.
First, \stream{a} has the value 1 and since no spawn condition is true both streams are not spawned.
Next, \stream{a} gets the value 4 spawning \stream{x} but not \stream{y}.
The later stream is spawned half a second later, which results in the two periods of \stream{x} and \stream{y} not being synchronized.
This leads to the failure of the synchronous access.
To circumvent this problem, the semantic type system requires equality instead of the implication of the spawn and close condition of two dependent periodic streams.

\paragraph{Type System Decidability}
Pacing types are defined as positive boolean formulas over input stream names or as fixed frequencies.
Hence, it is efficiently decidable if their implication is a tautology.
On the contrary, semantic types are arbitrary stream expressions.
As described by Schwenger~\cite{Schwenger_2022}, whether an implication between stream expressions is a tautology is generally undecidable.
Nevertheless, the \rtlola frontend provides a sound over-approximating implementation of the semantic type checker based on syntactic equality.
Here, semantic types are parsed as a boolean formula, so implications such as \lstinline{a.hold(or: 0) $\ \land$ b $\rightarrow$ a.hold(or: 0)} can still be proven.

\end{expertSection}

\section{Parameterization}
\label{sec:params}
The specifications in the previous sections were limited to a single intruder.
However, in reality, the monitor must observe an unbounded amount of intruders since we can not give an apriori bound on their number.
This monitor will inevitably require unbounded memory,
yet keeping the memory footprint of the monitor predictable is essential for CPS.
For this, \rtlola features parameterized output streams~\cite{DBLP:Lola2} that provide a declarative and predictable way of handling unbounded memory through stream expressions.

Parameterized streams lift output streams from a single instance to a set of stream instances.
While all stream instances share the stream expressions, each instance of a stream has a different assignment of parameter values.
This assignment is determined by an additional stream expression preceded by the \lstinline{with} keyword in the \lstinline{spawn} clause.
If an output stream is parameterized over multiple parameters, this expression returns a tuple of values matched position-wise to the parameters.
Parameters are declared as a comma-separated sequence in braces after the stream name.
The spawn clause of a parameterized stream determines when an instance is created as introduced in \Cref{sec:lifecycle}.
The evaluation and close clauses of parameterized streams are evaluated for each instance, determining their value and lifecycle.
As for dynamic streams, a stream instance will not be spawned again if an instance with the parameter values already exists.

Consider the final iteration of the running example which extends each output stream with a parameter for the different intruders:
\begin{example}[\spec{Multiple Intruder}{https://rtlola.cispa.de/playground/?spec=G-oEQIzEOEbxhyhtk9-33z5DSRefRXcxT3Y6-bkN9iVArNrUWKxAgXR39yxVMWhUV5C4uFplJPHVx5eqRfx93sLAw1hX_cp1w03g12Lo4ilPWOQWYoO3lzgzmGaUpX8JUi6ftlNqgb193FMW9PQ5uxRq9VrWvVXzrKLAFqxOAKhy451OeHQZFS6NdFRfG4GNQwsR2QLBmR4iUAa4DrrnmImusmIjdgDQK9aIsvLeiwWA1Qr8qHDunIkAbne7kVIUwOOu7xOLCGJHLFysipKJwgW8iRjLcVZ_VsTZVo6lwAixwQ4x9Xqi35-OJQZQpUOZZunYwsliqyoPYnYnQf6GjcvmdztSAnp304bA8UAaMXAs3GhU-56zktWFkKOgu9RB27dKVsNHCeL5yKMKHCRY6lqqTvQGSqIIBBB1q3FVxg_YQ46VX-HtRyHy2NujQPtcOhsLx1x3nJqOc2UhF4NR876-xe_v_z9shx2LHbs5rnNjyB4=&trace_name=c2ltcGxlX3RyYWNl&trace=G3cjEZW6FkAjITdletctxDHyxTy94e-6d_5LcahFCTcz3LusXT8XFUGAf97lN5sEusKWJosqTXokrjbNQWERCtwiFULmz7tvywS63OIQGjvJViIkn-jXR7N6pt8unoKCmDhcDOQ3x-WHZ_z8BDT94H3hf9rgsx-5rOvOb1j4-snHnW-xcD34NXcP-5tsNm8seCTzFH6bN-zMuG9ahf0GLxBNRTcsffdmmSXWlUK-_8e39ZZSfWtMucmjsynOxcLWrlA9PZnVXKs3d3al0FfnXvl-EqDD7m3w9vfl3_hm_bLx9pwM4KmMtLnou4tr0uHZqjkVRll_oMDyTbaDHX_9nPFs3-QgjJ8PbnaztSA9Eurdy8666bkRaoWiOtPRg9kUoaK-8wrYCsN1WFMkbmgeQFWqk1HlMFTA8cY7Zy6QFRe7UxikNvxmL85m7aA2GQFvNBeBCxetHsBDkoOlVmW0TW9h5Y4HewI6V03D6hWXCy-NXhkBFl6NyJK3BXmFVlCnsVrceUZLdPeHZ2HKd3XsEJdHtdUkxrVLxV2lUjcMt1wEB1JWHRIeqXZthO2IuepD3bg3rGaXSWFacZnpnhJzMQvupplc64QSDYvMfPucux1WvJ0O3Osltg6theSubjs4kl05zfTpaLpncrs56Cyws4TdTWhXAyoL1WUWXtbuuoZ7CAxi7w6t7b3X2dMeoLN1H5_F3vOkyLM4xXHmWKxKdqaoZOji6rqYWsgexSHq2EFxE9SxalSeZEgcvub1uTaL4XPDrrBjLBrgFlcGMlIoRc2w5xM73VXJZdbI5rRZmzUkaJo6-HqhEyr1foxCpu8hlqVbMOmWbDPSppWIPZpXwkjunoNsbOq8vWvIBVnBQ06TuZm1senBrDgjVO6kcI0bolOs-YssT6MTJF37Qkoc39FeEmInFEzw5nD52OHORf0dSmEAbMfXuy7UH9hphSStBZ6QHomZ7lwdsZk-bk9ErHiCOwN10WyfasiO61rGKIuOJTDsQnWtjLRwGuZ_rcvGwoh7_Py6oGBW3l5gV6pixtfevilPVLfGoqS16pi2zlH1bIHWfn6pzvTczJ0xt-s-pMlDKE9Z6zuE9rW3apO0Ta-kjsPBVoR0rkK7WPBymp1i9zDDMqXqEbbAGGmKe-d1OTZSAc64OlfjpM5XN8NLqCKDnEXV3A0EZ0dNsosu71VVTbmuLkN7S2LXyA5etOv6kI7OcXTvTvMwQ6mGi2eMD0vXq4knaY-T6qCB9ja86Xppjb5tAFW-baW7uqmZRThVpQdHuogetL3dIO8sSxSuTwrX-sShzymFYvooprYMwJXTLEdBroftMwrZ6HUvBrnZ-7peCh2vS6W6xK1ahgom5bvryOMo3K5GI3Fdp-4acVcvTy110xx6QNY2eWaVNlBYEJXu3UMpsshdNTbIldPcYdW6beqmN8fekSLjLLYRyZsJ7gxua7ZqgprtTrDyWG-2K3Qdt7_u5ytUR-3YtWy5DEeMTN5ae2j3KAVdboHxarD-gX287mltZSfM5mgVK2poDx3TFZ9ug7qpslfYji5UV-WGuLtnr1KFEY6Zl_7zuvYQiTG7DvpcAaLpZdlVlqISqAShbJjy23KkAcaRBk9GWQhHbOU6SZydEgH1gMNuaEGxxZFOPcB4NrlrC-iMjvbOqeoVy1XVVE_PZWiS2js0rXAhIDC4DSPRvtbM5XjQd26GcoE9PNg3PDlu91LA0dye22bOmjfNsrdsV82QxuFKUY1eO4i42G3aw5_NF_OqsKXtmQ0JVb2HArcO37lQPOzWzeDWVXHm4t1fS-01CddsamLdi5Z9Hg7MTB1nezF5l9qFB6Ou11uDw3UzaTfjkD1gx4GTGKVqYSDdI4-Lu3TtDHdqujTJtxJdLvBg0m7KdRVu0Lu0mG4bMi4LCKXn6kAjKnSI8WnyQW2GLosWdnPstutUnTF3aWm2Tgmk6ClodrO5prhzjZUVQhl31RDXs-RmuoAsVHnsM7BL082a6fKB0RsvqzAUzj905Lva-CBLbB53loXzDlM9k8Gal5pd2jc7UTAx6yFUaJjbaopjyDl_zdelU1zVVvPcXOywXYbJiSlLa1jZ6cxpT1bemjq6J5feryngNsvdpxP3sCYrcOCF5-7sCecZDJbcU5Uvs-zyaFBFX-fzda691MhoFiDUyqzMzfFC7uzSKmgzCaatysS6V7ntHdV2Bo5qpG7AvRZ6PPEGV4z7KhybXeztsqpuabmOr2rIVHedqM10zSx7q47hreC0EmhsAahOWVWl43QfbG66wdEVAZlf6ynVlZuc5A3daNOp3Y70ce681LbC1tQonKnpSZ69NbmaGlCUHc71zOaoYBFxjsAN3OvYi026emAXaibAaXOLwqG0FNM7cW7La26PWKQwBaesLat560XcnMJRN62cCmtzc4MlprQnfX2vbhlpsXs-WWwrblmJBqpXaZyaRrkb7r5w5afEzO2x7MRr2l1CZ-c2m2emDmzO2sHcZIRqZHD3AMwxd5LddsKxxBZP4rADnkD6yuS2DaxROqoLzallB8JnzprEpU5Ajrk9Efr7FWZ45apzD2Ftq3TcBCwwOqdXMV1JFZlCfGbTfe1wroqJ3hxyWTckuePmPk0XrqlINLvJwWq2m2vJJ5twxdnubcFPOI1PsZTFwusJ2gPMPCesq77a7g3CU2yVs8qukpI4BFL42hYJvLVlzLiFlWLTQWDtzO0No7loCm1Ob-qIWbIXVd1i2vYsqFEMLYaWvuJYID3czYjblVWQ2mGv1orNy0JIDpO-fJ3XFex5atxxFlX60NcpGNS22GqeEFFzGuyA5gQp9E7VMeDXeVL3VRUvxpg-wJyUhVvxyGd4N4yq1Z25W3ZJ7M2VDs-k2BxUrlZk2CjsdYZNPA5rvFelvXVns9lLMjR952ovt6yxo1Otb33X0q2Ls97xOSvDPkZZDMGl-7ZqttOESFRfHaaUOajz8aoHullzHygB5AxdQyLrrh6uTN201pBwGatdN5WBUpvpLKFuczVL6KeDVz6Va7qWHSl9cXauKqOFm8r0PI2OcwUA85uiTVXYMFlvg_SVcgSFHkk1raseq04lZ2rbBkN7R9bN6Zm41lwEVdzwGEZLwoVkZbOT6lvTPLgHGPdN8q2cij2ZYmJvijJWk1d1cUB3FpMWtA3B8c1GszGzR7Exk-c1oqQnJbHbbmex3NtZDO7sFO7KPWHczBkOOwxvupNxUbnJUrS8jGOp3SHXQcoHORBiZm9wK8LqVCgOssjxDmxN3bTB2sIuWv95BVhAqZbEsFuXS40wByuZgtEER0-g5u64CpMtcKq1mkbStD5ci8suWopJmOrbRTNeo48n2rW3RRcpmSlGO_HsjDTND_FeGzOG57ma5rkQ3EvFoD0O0qHufBw0Z0GLwM3GcZyNNbpK3Uw3L510qeIbd-pS3XudmdrbPXVd8SlznaKI41u3t7h24YY7qSWPJVihk7mdqDO11Wt5Z9NVOgWntKntm0kftVl3YOFcUJyIY6ZYRx6X05nuWdrl7FRL09f8Ld3hynsW0VZ9mI5cZ2Fi0lNj4cTRV2SlY1o-nrFXtczsaxVA04u9I1i355UKyN58QVisduLRxZ1RO4fdGe5zD3o2XuYfKkmFwHpvhe1c5WvRGHaIqCnuQT2bvXPOzWyJwUXLzc2K8nc77ngOgribku31LnuyexE0qkGooXxTNaBDEjeXmuKdjBvNREF1zG92mu9d2pr1GaY8dgUdwqtezwCDgZdqR9vbnYv2KtzXnh3XcpzdarJQK10dtGQBVG313rgvDQO7F9zVTZ3P2iysswV2BpZdLsyMBJuT7DArL3fqXEUnRleXq1sGI3OqlY8b14j8hq9Wc00NT_eiG7ebnC4x06d1tE2LTZGVYm6eJJfZ2atMV76rCgjhFKJW620WuF7o5lCTWbZgyPVEbSvZOxU4Z6MFfmWHGh33VLdbLk7duVHMbsiSwLoaVVXyqByv6tO4p1czLahjlXqGTncDuG5RJQpYUHJFqSlNC6CXclqQTLPP6wzPnpM6XyS2t6YSPdzq5jyCp0jVlY--E0zLx5ux1FyEmh4Z19560jXO6mEbXRkW6lqXC3A7HHLoZJXpve2YdKWqUSJ8ztd8R26fB5dm7HT2Tsxmqe4oHRwPFz3rDkNj1wmalrnLRLqPUA11-R9O}]
	\label{ex:full}
	\begin{lstlisting}
	import math
	input lat: Float
	input lon: Float
	input intruder_id: UInt
	input intruder_lat: Float
	input intruder_lon: Float
	
	output intruder_pos(id)
	    spawn with intruder_id
	    eval when id = intruder_id with  (intruder_lat, intruder_lon)
	    close @true@ when stale(id).hold(or: false)  
	output distance(id)
	    spawn with intruder_id
	    eval @((intruder_id && intruder_lat && intruder_lon) || (lat &&lon))@
	    with sqrt((intruder_pos(id).hold().0.defaults(to: 0.0) - lat.hold(or: 0.0))**2.0 + (intruder_pos(id).hold().1.defaults(to: 0.0) - lon.hold(or: 0.0))**2.0)
	    close @true@ when stale(id).hold(or: false)  
	output closer(id)
	    spawn with intruder_id
	    eval with distance(id).offset(by: -1).defaults(to: distance(id)) >= distance(id)
	    close @true@ when stale(id).hold(or: false) 
	output stale(id)
	    spawn with intruder_id
	    eval @10s@ with intruder_pos(id).aggregate(over: 10s, using: count) = 0
	    close @true@ when stale(id).hold(or: false)
	
	trigger(id)
	    spawn when distance(intruder_id).hold(or: 1.0) < 0.1 with intruder_id
	    eval @1Hz@ when closer(id).aggregate(over_exactly: 5s, using: forall).defaults(to: false) with "Intruder {{}} detected".format(id)
	    close @true@ when stale(id).hold(or: false) 
	\end{lstlisting}
\end{example}
Notice the additional input stream \stream{intruder\_id}.
We assume that every intruder has a unique ID provided to the monitor with every update of the \stream{intruder\_lat} and \stream{intruder\_lon} streams.
The output stream \stream{intruder\_pos} is added to accumulate these positions on a per-intruder basis.
It has a single parameter representing the intruder ID.
This is made explicit through its \lstinline{spawn with} expression that synchronously accesses the \stream{intruder\_id} stream.
As a result, a new instance of this stream is created for each fresh intruder when it is received for the first time.
All other output streams share the same parameterization, so each output stream has an instance for each non-stale intruder ID.
The trigger features the additional \lstinline{spawn when} condition as before.
Finally, this specification achieves the goal outlined in \Cref{sec:overview} and handles multiple moving intruders efficiently and predictably.

\begin{expertSection}{Instance Aggregations}
Like sliding window aggregations, \rtlola features instance aggregations to aggregate the most recent values of the instances of a parameterized output stream.
For example, the trigger from \Cref{ex:full} can be rewritten without parameters using an instance aggregation \spec{as follows}{https://rtlola.cispa.de/playground/?spec=Gy4FABwHdtPOg_fWGCHlYpv_qdtclunh-dUEvD5xDAuzaWoaSQHyP5cTuKlwdIf2x61Rm6K1OJrbuqzn0_MwhsytffwYFRSbztuWQ_AhRb4yioL3PMFK3yCvpX2QybWk6iXxX75xmVePZ37-PuxTzLQkTTnGRKoDJRnPg2lSsO-DmMMAUKTXJp7yGGsW-TDU5bVKUg94ZajSGRigdR0DN4GR0A2tRtJaRXmVTwjwkTXySf70VQhAa5p-UUhTIWMQz2czWzHNAoid32dMIHFL2MbbIstLxgxmIa6SyU2__ixYmdjc5AQD8lqqhPTPkaouTA4NwjQIwyRemkGx4oHyHdnjbqD_YeP9rbMNQeDHC98QfI-4YOhWwmKL2qqh4MU50ejWK51CF03s0p4-l-jZM5IPJXgc7mjnMoPM1W_8TijHsooFb1ymK9x_h8tNoaRw7Nmp09QLzrHzwl5r-faV5FcpHcdqLNxDZQB3HCGp4pgJJoIaW1ijb1-lzBTBq4SRugN4PMvn_ZJ5jAA=&trace_name=c2ltcGxlX3RyYWNl&trace=G3cjEZW6FkAjITdletctxDHyxTy94e-6d_5LcahFCTcz3LusXT8XFUGAf97lN5sEusKWJosqTXokrjbNQWERCtwiFULmz7tvywS63OIQGjvJViIkn-jXR7N6pt8unoKCmDhcDOQ3x-WHZ_z8BDT94H3hf9rgsx-5rOvOb1j4-snHnW-xcD34NXcP-5tsNm8seCTzFH6bN-zMuG9ahf0GLxBNRTcsffdmmSXWlUK-_8e39ZZSfWtMucmjsynOxcLWrlA9PZnVXKs3d3al0FfnXvl-EqDD7m3w9vfl3_hm_bLx9pwM4KmMtLnou4tr0uHZqjkVRll_oMDyTbaDHX_9nPFs3-QgjJ8PbnaztSA9Eurdy8666bkRaoWiOtPRg9kUoaK-8wrYCsN1WFMkbmgeQFWqk1HlMFTA8cY7Zy6QFRe7UxikNvxmL85m7aA2GQFvNBeBCxetHsBDkoOlVmW0TW9h5Y4HewI6V03D6hWXCy-NXhkBFl6NyJK3BXmFVlCnsVrceUZLdPeHZ2HKd3XsEJdHtdUkxrVLxV2lUjcMt1wEB1JWHRIeqXZthO2IuepD3bg3rGaXSWFacZnpnhJzMQvupplc64QSDYvMfPucux1WvJ0O3Osltg6theSubjs4kl05zfTpaLpncrs56Cyws4TdTWhXAyoL1WUWXtbuuoZ7CAxi7w6t7b3X2dMeoLN1H5_F3vOkyLM4xXHmWKxKdqaoZOji6rqYWsgexSHq2EFxE9SxalSeZEgcvub1uTaL4XPDrrBjLBrgFlcGMlIoRc2w5xM73VXJZdbI5rRZmzUkaJo6-HqhEyr1foxCpu8hlqVbMOmWbDPSppWIPZpXwkjunoNsbOq8vWvIBVnBQ06TuZm1senBrDgjVO6kcI0bolOs-YssT6MTJF37Qkoc39FeEmInFEzw5nD52OHORf0dSmEAbMfXuy7UH9hphSStBZ6QHomZ7lwdsZk-bk9ErHiCOwN10WyfasiO61rGKIuOJTDsQnWtjLRwGuZ_rcvGwoh7_Py6oGBW3l5gV6pixtfevilPVLfGoqS16pi2zlH1bIHWfn6pzvTczJ0xt-s-pMlDKE9Z6zuE9rW3apO0Ta-kjsPBVoR0rkK7WPBymp1i9zDDMqXqEbbAGGmKe-d1OTZSAc64OlfjpM5XN8NLqCKDnEXV3A0EZ0dNsosu71VVTbmuLkN7S2LXyA5etOv6kI7OcXTvTvMwQ6mGi2eMD0vXq4knaY-T6qCB9ja86Xppjb5tAFW-baW7uqmZRThVpQdHuogetL3dIO8sSxSuTwrX-sShzymFYvooprYMwJXTLEdBroftMwrZ6HUvBrnZ-7peCh2vS6W6xK1ahgom5bvryOMo3K5GI3Fdp-4acVcvTy110xx6QNY2eWaVNlBYEJXu3UMpsshdNTbIldPcYdW6beqmN8fekSLjLLYRyZsJ7gxua7ZqgprtTrDyWG-2K3Qdt7_u5ytUR-3YtWy5DEeMTN5ae2j3KAVdboHxarD-gX287mltZSfM5mgVK2poDx3TFZ9ug7qpslfYji5UV-WGuLtnr1KFEY6Zl_7zuvYQiTG7DvpcAaLpZdlVlqISqAShbJjy23KkAcaRBk9GWQhHbOU6SZydEgH1gMNuaEGxxZFOPcB4NrlrC-iMjvbOqeoVy1XVVE_PZWiS2js0rXAhIDC4DSPRvtbM5XjQd26GcoE9PNg3PDlu91LA0dye22bOmjfNsrdsV82QxuFKUY1eO4i42G3aw5_NF_OqsKXtmQ0JVb2HArcO37lQPOzWzeDWVXHm4t1fS-01CddsamLdi5Z9Hg7MTB1nezF5l9qFB6Ou11uDw3UzaTfjkD1gx4GTGKVqYSDdI4-Lu3TtDHdqujTJtxJdLvBg0m7KdRVu0Lu0mG4bMi4LCKXn6kAjKnSI8WnyQW2GLosWdnPstutUnTF3aWm2Tgmk6ClodrO5prhzjZUVQhl31RDXs-RmuoAsVHnsM7BL082a6fKB0RsvqzAUzj905Lva-CBLbB53loXzDlM9k8Gal5pd2jc7UTAx6yFUaJjbaopjyDl_zdelU1zVVvPcXOywXYbJiSlLa1jZ6cxpT1bemjq6J5feryngNsvdpxP3sCYrcOCF5-7sCecZDJbcU5Uvs-zyaFBFX-fzda691MhoFiDUyqzMzfFC7uzSKmgzCaatysS6V7ntHdV2Bo5qpG7AvRZ6PPEGV4z7KhybXeztsqpuabmOr2rIVHedqM10zSx7q47hreC0EmhsAahOWVWl43QfbG66wdEVAZlf6ynVlZuc5A3daNOp3Y70ce681LbC1tQonKnpSZ69NbmaGlCUHc71zOaoYBFxjsAN3OvYi026emAXaibAaXOLwqG0FNM7cW7La26PWKQwBaesLat560XcnMJRN62cCmtzc4MlprQnfX2vbhlpsXs-WWwrblmJBqpXaZyaRrkb7r5w5afEzO2x7MRr2l1CZ-c2m2emDmzO2sHcZIRqZHD3AMwxd5LddsKxxBZP4rADnkD6yuS2DaxROqoLzallB8JnzprEpU5Ajrk9Efr7FWZ45apzD2Ftq3TcBCwwOqdXMV1JFZlCfGbTfe1wroqJ3hxyWTckuePmPk0XrqlINLvJwWq2m2vJJ5twxdnubcFPOI1PsZTFwusJ2gPMPCesq77a7g3CU2yVs8qukpI4BFL42hYJvLVlzLiFlWLTQWDtzO0No7loCm1Ob-qIWbIXVd1i2vYsqFEMLYaWvuJYID3czYjblVWQ2mGv1orNy0JIDpO-fJ3XFex5atxxFlX60NcpGNS22GqeEFFzGuyA5gQp9E7VMeDXeVL3VRUvxpg-wJyUhVvxyGd4N4yq1Z25W3ZJ7M2VDs-k2BxUrlZk2CjsdYZNPA5rvFelvXVns9lLMjR952ovt6yxo1Otb33X0q2Ls97xOSvDPkZZDMGl-7ZqttOESFRfHaaUOajz8aoHullzHygB5AxdQyLrrh6uTN201pBwGatdN5WBUpvpLKFuczVL6KeDVz6Va7qWHSl9cXauKqOFm8r0PI2OcwUA85uiTVXYMFlvg_SVcgSFHkk1raseq04lZ2rbBkN7R9bN6Zm41lwEVdzwGEZLwoVkZbOT6lvTPLgHGPdN8q2cij2ZYmJvijJWk1d1cUB3FpMWtA3B8c1GszGzR7Exk-c1oqQnJbHbbmex3NtZDO7sFO7KPWHczBkOOwxvupNxUbnJUrS8jGOp3SHXQcoHORBiZm9wK8LqVCgOssjxDmxN3bTB2sIuWv95BVhAqZbEsFuXS40wByuZgtEER0-g5u64CpMtcKq1mkbStD5ci8suWopJmOrbRTNeo48n2rW3RRcpmSlGO_HsjDTND_FeGzOG57ma5rkQ3EvFoD0O0qHufBw0Z0GLwM3GcZyNNbpK3Uw3L510qeIbd-pS3XudmdrbPXVd8SlznaKI41u3t7h24YY7qSWPJVihk7mdqDO11Wt5Z9NVOgWntKntm0kftVl3YOFcUJyIY6ZYRx6X05nuWdrl7FRL09f8Ld3hynsW0VZ9mI5cZ2Fi0lNj4cTRV2SlY1o-nrFXtczsaxVA04u9I1i355UKyN58QVisduLRxZ1RO4fdGe5zD3o2XuYfKkmFwHpvhe1c5WvRGHaIqCnuQT2bvXPOzWyJwUXLzc2K8nc77ngOgribku31LnuyexE0qkGooXxTNaBDEjeXmuKdjBvNREF1zG92mu9d2pr1GaY8dgUdwqtezwCDgZdqR9vbnYv2KtzXnh3XcpzdarJQK10dtGQBVG313rgvDQO7F9zVTZ3P2iysswV2BpZdLsyMBJuT7DArL3fqXEUnRleXq1sGI3OqlY8b14j8hq9Wc00NT_eiG7ebnC4x06d1tE2LTZGVYm6eJJfZ2atMV76rCgjhFKJW620WuF7o5lCTWbZgyPVEbSvZOxU4Z6MFfmWHGh33VLdbLk7duVHMbsiSwLoaVVXyqByv6tO4p1czLahjlXqGTncDuG5RJQpYUHJFqSlNC6CXclqQTLPP6wzPnpM6XyS2t6YSPdzq5jyCp0jVlY--E0zLx5ux1FyEmh4Z19560jXO6mEbXRkW6lqXC3A7HHLoZJXpve2YdKWqUSJ8ztd8R26fB5dm7HT2Tsxmqe4oHRwPFz3rDkNj1wmalrnLRLqPUA11-R9O}:
\begin{lstlisting}
	output approaching(id)
	    spawn when distance(intruder_id).hold(or: 1.0) < 0.1 with intruder_id
	    eval @1Hz@ with closer(id).aggregate(over_exactly: 5s, using: forall).defaults(to: false)
	    close @true@ when stale(id).hold(or: false) 
	    
	trigger @true@ approaching.aggregate(over_instances: all, using: exists) "Intruder detected"
\end{lstlisting}
The \stream{approaching} stream is similar to the trigger from \Cref{ex:full}.
The non-parameterized trigger now aggregates over all instances of this new stream using a disjunction.
Semantically, this means that whenever any instance of the helper stream evaluates to true, the instance aggregation in the trigger will also evaluate to true, causing the trigger to activate.

Instead of aggregating all stream instances, it is sometimes desirable to only aggregate the instances that produced a \emph{fresh} value in this evaluation cycle.
This is specified by stating \lstinline{over_instances: fresh} in the aggregation.
This will couple the pacing of the caller of the aggregation to the evaluation pacing of the target stream of the aggregation.
Other instance aggregation functions in \rtlola are \lstinline{forall, avg, sum, count, min}, and \lstinline{max}.


\end{expertSection}
\section{Development and Integration}
\label{sec:evaluation}
During the previous sections, the examples featured links to the \rtlola playground, a web-based implementation of the \rtlola frontend and interpreter that can analyse and execute specifications locally in your browser without requiring installation.
A version of this tutorial is also available there\footnote{\url{https://rtlola.org/playground/tutorial}}, such that the specifications and examples from this tutorial can easily be experimented with by pressing the \emph{Copy to Editor} button below them and clicking \emph{Run}.
The specifications are tested against a trace simulated using the \emph{Microsoft Flight Simulator}.
A video of this trace is included in the overview chapter in the playground, and its raw data can be inspected by switching to the \emph{Trace} tab on the right.

Nonetheless, in real applications, it is necessary to run the monitor either natively or incorporate them within existing applications.
The \rtlola interpreter provides solutions for both:
A library to seamlessly integrate the monitor into Rust applications, along with a standalone command-line application.

\subsection{\rtlola CLI}

The simplest way to run the \rtlola interpreter locally is the command-line application \lstinline{rtlola-cli}.
The installation of the application can be achieved using the cargo package manager:
\begin{lstlisting}
	cargo install rtlola-cli
\end{lstlisting}
The application offers two modes of execution:
\begin{itemize}
	\item \emph{Analyze} In this mode, the \rtlola frontend is employed to check the provided specification for correctness. This involves checking for syntax errors, ensuring well-definedness and detecting type-related errors.
	\item \emph{Monitor} This mode enables the execution of a monitor based on the provided specification.
		After checking the specification for correctness, the monitor can be run in an offline or online setting.
		In an offline setting, the monitor analyzes a prerecorded trace, while in an online setting, the data arrives in real-time.
\end{itemize}

In the offline setting, the interpreter monitors a prerecorded trace in CSV format such as the following:
\begin{lstlisting}
	a,b,time
	1,2,0
	#,3,1
	3,#,2
\end{lstlisting}
This example trace defines three events occurring at times 0s, 1s, and 2s, respectively, and assigns new values to two input streams called \lstinline{a} and \lstinline{b}.
Notably, the hashmark \lstinline{#} signifies that the corresponding input stream does not receive a new value at that particular time.
Subsequently, we can run the \lstinline{rtlola-cli} to monitor the specification \texttt{specification.lola} on this trace:
\begin{lstlisting}
	rtlola-cli monitor \
		--offline relative \
		--csv-in trace.csv \
		specification.lola
\end{lstlisting}
Here, the argument \lstinline{--offline relative} specifies the type of time format utilized in the "time" column of the CSV file.
In the example, the timestamp is given as time in seconds relative to the start of the monitor.
Upon executing, each event in the CSV file is forwarded to the monitor, with the resulting output printed to the standard output.

In the online mode, the interpreter retrieves the inputs from a buffer and utilizes the real-time timestamps of the events when they arrive.
The following command starts the monitor in an online setting:
\begin{lstlisting}
	rtlola-cli monitor --online --stdin specification.lola
\end{lstlisting}
Here, the application waits for new events on standard input and forwards them to the monitor as soon as they arrive.

We illustrated two instances of using the \rtlola interpreter command line application.
For a comprehensive list of available command-line arguments and options, you can consult the documentation by executing the following command:
\begin{lstlisting}
	rtlola-cli monitor --help
\end{lstlisting}

The \rtlola interpreter has successfully been employed to monitor drones in cooperation with the German Aerospace Center (DLR) and the aircraft manufacturer Volocopter.
We identified a set of different monitoring applications, as reported by Baumeister et.al.~\cite{volostream}, which we discuss in the expert section.

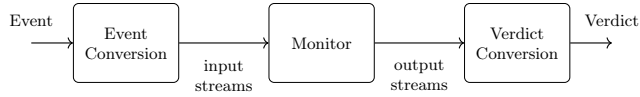
\begin{figure}[t]
	\centering
	\scalebox{0.7}{
	\begin{tikzpicture}[n/.style={draw,rectangle,minimum width=2cm,minimum height=1.5cm,align=center,rounded corners=1mm},p/.style={semithick}]
		\node[n] (event_conversion) {Event\\Conversion};
		\node[n,right=1.7cm of event_conversion] (monitor) {Monitor};
		\node[n,right=1.7cm of monitor] (verdict_conversion) {Verdict\\Conversion};
		\draw[<-,p] (event_conversion) -- ++(-18mm,0) node[above=2mm] {Event};
		\draw[->,p] (event_conversion) -- node[below=2mm,align=center] {input\\streams} (monitor);
		\draw[->,p] (monitor) -- node[below=2mm,align=center] {output\\streams} (verdict_conversion);
		\draw[->,p] (verdict_conversion) -- ++(18mm,0) node[above=2mm] {Verdict};
	\end{tikzpicture}}
	\caption{Process to integrate the monitor}
	\label{fig:integration}
\end{figure}

\begin{expertSection}{Monitoring Applications}
The monitor can provide valuable feedback during the development of new components in the aerospace domain.
In this safety-critical domain, pre-defined standards ensure that the concept of operation, requirements, design, and implementation are coherent and include several validation steps.
We identified different applications in which monitoring can be used during the development of new components following such standards but also during the operation of these components:
\begin{enumerate}
	\item \emph{Debugging} The monitor provides feedback to the developer of the component. During the execution, the monitor checks whether the component works as intended and collects statistical information. The developer writes the specification and has access to the internal state of the component.
	\item \emph{Validation} The monitor is used to validate the component externally. The specification is written independently and has only access to the inputs and outputs of the component and not to the internal state. This application is, among others, helpful in validating that components by external companies follow their requirements and can be trusted.
	\item \emph{Pre-Post-Flight Analysis} This application uses monitoring to check whether all necessary components are operational. The monitor runs pre-defined test cases and validates that no irregular behavior is detected.
	After the flight, the monitor computes more sophisticated information for better evaluation of the flight or runs new specifications on past flights.
%
	\item \emph{In-Flight Analysis / Safe Integration} The monitor provides feedback about the safety of the drone during its operation. It validates the correctness of individual components to ensure a safe flight and reports to the pilot if a property is violated.
\end{enumerate}

All these applications require the integration of the monitor into the development process.
Previously, we utilized the command-line application of the \rtlola interpreter to execute the monitor.
In the following example, we demonstrate how to integrate the interpreter using the Rust library.

\paragraph{Integration with the RTLola API}
We assume that the monitor is running on the drone and receives the input data over internal communication.
The output of the monitor is then sent over TCP to a ground station displaying the trigger messages of the monitor so a pilot can take over.
\Cref{fig:integration} illustrates two steps for this integration process.
As shown in the figure, the setup consists of three components:
The monitor, in the center of the figure, is automatically generated from the specification and does not require further integration.
However, two interfaces must be implemented to handle the communication with the system and the operator.
These implementations are specific to the setup:
The \emph{Event Conversion} receives the incoming sensor readings and transforms the data into an internal representation the monitor understands.
The \emph{Verdict Conversion} transforms the internal representation of the monitor's output to messages accepted by the ground station.

After providing all the missing implementations, we need to configure the monitor and start the evaluation.
This results in the following code, skipping the concrete configuration of the \lstinline!event_source! and the \lstinline!verdict_sink!:
\begin{lstlisting}[language=Rust, basicstyle=\ttfamily\scriptsize]
	let mut monitor = ConfigBuilder::new()
        .with_spec(spec)
        .online()
        .with_event_factory::<ExampleInputs::Factory>()
        .with_verdict::<TriggerMessages>()
        .monitor()?;
    let mut event_source = ...
    let mut verdict_sink = ...
    while let Some((ev, ts)) = event_source.next_event()? {
        let verdicts = monitor.accept_event(ev, ts)?;
        verdicts_sink.sink_verdicts(&verdicts)?;
    }
\end{lstlisting}


\paragraph{Event Conversion}
The Event Conversion receives sensor values and transforms the readings into an internal representation the monitor uses.
In our setup, we receive the sensor values over a UDP connection as a byte-stream and differentiate between two types of messages:
The first type of message is sent by the GNSS sensor that computes the latitude and longitude of the drone.
The second type of message contains the latitude and longitude of the intruders.
The byte-stream needs to be parsed and converted to map the incoming data to the corresponding input streams.
The implementation in our setup uses the simplified interfaces shown in \Cref{fig:concrete_event_conversion}.
After providing a parser function, the implementation receives the byte stream and parses this stream to an event called \lstinline{ExampleInputs}.
The implementation of the interface is provided automatically by the macros \lstinline{ValueFactory} and \lstinline{CompositeFactory}.
These macros generate code for a factory that maps, for example, the \lstinline!lat! field in the \lstinline!Gnss! struct to the \lstinline!lat! input stream in the monitor.

\paragraph{Verdict Conversion}
The Verdict Conversion transforms the internal representation of the monitor's output into messages in a form expected by the ground station.
In our setup, this conversion interprets trigger messages as bytes and sends these over TCP to the ground station.
The interfaces for this setup are implemented generically, so no further steps are needed.


\end{expertSection}

\begin{figure}[t]
	\begin{subfigure}{\textwidth}
		\begin{lstlisting}[language=Rust, basicstyle=\ttfamily\scriptsize]
		impl ByteParser for DroneExampleParser {
			fn from_bytes(&mut self, bytes: &[u8]) -> Result<(ExampleInputs, usize)> { ... }
		}
		\end{lstlisting}
	\end{subfigure}
	\begin{subfigure}{0.3\textwidth}
		\begin{lstlisting}[language=Rust, basicstyle=\ttfamily\scriptsize]
		#[derive(ValueFactory)]
		struct Gnss {
			lat: Float64,
			lon: Float64
		}
		\end{lstlisting}
	\end{subfigure}
	\begin{subfigure}{0.3\textwidth}
		\begin{lstlisting}[language=Rust, basicstyle=\ttfamily\scriptsize]
		#[derive(ValueFactory)]
		#[factory(prefix)]
		struct Intruder {
			id: UInt64
			lat: Float64,
			lon: Float64
		}
		\end{lstlisting}
	\end{subfigure}
	\begin{subfigure}{0.3\textwidth}
		\begin{lstlisting}[language=Rust, basicstyle=\ttfamily\scriptsize]
		#[derive(CompositFactory)]
		enum ExampleInputs {
			Gnss(Gnss),
			Intruder(Intruder)
		}
		\end{lstlisting}
	\end{subfigure}
	\caption{Concrete Implementation of the Event Conversion}
	\label{fig:concrete_event_conversion}
\end{figure}

\section{Conclusion}
\label{sec:conclusion}

The running example from our tutorial has illustrated the expressiveness
of the \rtlola specification language, which makes \rtlola well-suited for
complex application domains like aerospace. The development of the
specifications is facilitated by the comprehensive support of the tool
framework. Since the same specification can be used in multiple
different settings, a specification can be validated early in a test
environment or on log data, long before the monitor is integrated into
the aircraft; the automatic analysis of the specification furthermore
ensures that the monitor operates correctly and reliably.

\rtlola has been very successful in the aerospace domain
(cf.~\cite{DBLP:ClearedForTakeOff,volostream}). \rtlola has also been
used in other cyber-physical applications, including
cars~\cite{DBLP:journals/sttt/BiewerFHKSS23} and medical
equipment~\cite{10.1145/3446913.3460318}, and in domains beyond CPS,
such as networks~\cite{DBLP:Lola2}.  The combination of the highly
expressive \rtlola specification language with the reliability obtained by
static analysis and the resource efficiency of the monitoring
framework is of great use in all these settings.






\subsubsection*{Data Availability Statement}

The artifacts and resources associated with this paper are accessible as follows:

\begin{enumerate}
    \item \textbf{Primary Artifact:} The primary artifact of this paper is available via:\\\url{https://doi.org/10.5281/zenodo.12633784}.
    
    \item \textbf{Source Code:} The source code of the framework is hosted on GitHub:

      \vspace{1mm}\begin{tabular}{ll}
        RTLola Interpreter: &\url{https://github.com/reactive-systems/RTLola-Interpreter}\\
        RTLola Frontend: &\url{https://github.com/reactive-systems/RTLola-Frontend}
      \end{tabular}
    
    \item \textbf{Software Packages:} The relevant software packages are available on crates.io:
    \vspace{1mm}\begin{tabular}{ll}
        RTLola Interpreter: &\url{https://crates.io/crates/rtlola-interpreter}\\
        RTLola Frontend: &\url{https://crates.io/crates/rtlola-frontend}
    \end{tabular}
\end{enumerate}

\bibliographystyle{splncs04}
\bibliography{references}

\end{document}